\documentclass[twocolumn,prl,superscriptaddress,floatfix,showpacs]{revtex4-1}
\usepackage[utf8]{inputenc}
\usepackage{amsmath}
\usepackage{amssymb}
\usepackage{graphicx}

\makeatletter
\usepackage{graphics}
\usepackage{epsfig}
\usepackage{color}

\newcommand{\be}{\begin{equation}} \newcommand{\ee}{\end{equation}}
\newcommand{\bea}{\begin{eqnarray}} \newcommand{\eea}{\end{eqnarray}}

\usepackage[caption=false]{subfig} 
\usepackage[caption=false]{subfig} 
\usepackage[usenames,dvipsnames,svgnames,table]{xcolor}
\definecolor{lightyellow}{cmyk}{0,0,0.3,0}
\definecolor{lightblue}{cmyk}{0.1,0,0,0}
\definecolor{green4}{cmyk}{0.0,0.0,0.7,0.4}
\definecolor{dgreen}{rgb}{0,.4,0}
\definecolor{yellow2}{rgb}{.7,.7,0}
\definecolor{indianred}{rgb}{0.8,0.36,0.36}

\makeatother

\begin{document}

\title{Immunization and targeted destruction of networks using explosive percolation}

\author{Pau Clusella} \affiliation{Institute for Complex Systems and Mathematical Biology, SUPA, University of Aberdeen, Aberdeen AB24 3UE, United Kingdom} 
\affiliation{Dipartimento di Fisica e Astronomia, Università di 
Firenze, via G. Sansone 1, I-50019 Sesto Fiorentino, Italy}
\author{Peter Grassberger} \affiliation{JSC, FZ J\"ulich, D-52425 J\"ulich, Germany}\affiliation{Institute for Complex Systems and Mathematical Biology, SUPA, University of Aberdeen, Aberdeen, UK}
\author{Francisco J. P\'erez-Reche} \affiliation{Institute for Complex Systems and Mathematical Biology, SUPA, University of Aberdeen, Aberdeen AB24 3UE, United Kingdom}
\author{Antonio Politi} \affiliation{Institute for Complex Systems and Mathematical Biology, SUPA, University of Aberdeen, Aberdeen AB24 3UE, United Kingdom}

\date{\today}

\begin{abstract}
A new method (`explosive immunization' (EI)) is proposed for immunization and targeted destruction of networks. It
combines the explosive percolation (EP) paradigm with the idea of maintaining a fragmented distribution
of clusters. The ability of each node to block the spread of an infection (or to prevent the existence 
of a large cluster of connected nodes) is estimated by a score. The algorithm proceeds by first identifying 
{\it low score} nodes that should not be
vaccinated/destroyed, analogously to the links selected in EP if they do {\it not} 
lead to large clusters. As in EP, this is done by selecting the worst node (weakest blocker) from a finite set of randomly chosen 
`candidates'. 
Tests on several real-world and model networks suggest that the method is more efficient and faster than 
any existing immunization strategy. Due to the latter property it can deal with very large networks.
\end{abstract}
\maketitle

Network robustness is a major theme in complex-systems theory that has attracted much attention in recent years 
\cite{Newman_Book2010}.
Two specific problems are immunization of networks against epidemic spreading (of infection diseases, computer viruses, 
or malicious rumors), and the destruction of networks by targeted attacks. At first sight these two look completely different, 
but they can actually be mapped onto each other. 
The key observation is that infection spreading in a population use the network of contacts between hosts for their spread. 
Accordingly, from the viewpoint of the infection, immunization corresponds to an attack that destroys the network on which 
it can spread. Vaccination of hosts (network nodes) is often the most effective  way to prevent large epidemics.  
Other strategies include manipulating the network 
topology \cite{Schneider_PRE2011,Schneider_SciRep2013,Zeng-Liu_PRE2012} or introducing heterogeneity 
in transmission of the infection~\cite{PerezReche_JRSInterface2010,Neri_JRSInterface2010,Neri_PLoSCBio2011}. 

The main task in both cases is to find those nodes (``blockers")
whose removal is most efficient in destroying connectivity. Important blockers (``superblockers'') are often assumed
\cite{Morone_Nature2015} to be equivalent to ``superspreaders", i.e. the most efficient nodes in
spreading information, supplies, marketing strategies, or technological innovations. Identifying superspreaders is the subject 
of a vast literature
\cite{Newman_Book2010} but, as pointed out in, e.g., Ref.~\cite{Habiba_Chapter2010}, identifying superblockers is not 
the same as finding superspreaders. Indeed, a node in a densely connected core will in general be a good 
spreader~\cite{Pei_JSTAT2013}, but it will be in general a very poor blocker, since the infection can easily 
find ways to go around it.

Here we devise a strategy which identifies superblockers. Vaccinating such nodes provides an efficient way to fragment 
the network and reduce the possibility of large epidemic outbreaks. We focus on ``static'' immunization which aims at 
fragmenting the network before a possible outbreak occurs (``dynamical'' immunization strategies where one tries to 
contain an ongoing epidemic were studied, for instance, in \cite{Wu_PhysicaA2015_DynamicImmunization}). In our approach, 
the network consists of $N$ nodes out of which $qN$ are vaccinated; the rest are left susceptible to the infection. 
The size of an invasion will depend on the fraction $q$ of vaccinated nodes, the type of epidemic (e.g. Susceptible-Infected-Removed or 
Susceptible-Infected-Susceptible~\cite{Hebert-Dufresne_SciRep2013}), and its virulence. However, the maximum fraction of nodes infected at any time 
will always be bounded by the relative size $S(q)$ of the largest \emph{cluster of susceptible} nodes, ${\cal G}(q)$.
Keeping $S(q)$ as small as possible will therefore ensure that epidemic outbreaks of any type are as small as they can 
be for a vaccination level $q$~\cite{Morone_Nature2015,Schneider_EPL2012}. For large 
networks, $N\rightarrow \infty$, the aim of immunization is to fragment them so that $S(q)=0$~\cite{Morone_Nature2015}.
The immunization threshold $q_c$ is defined as the smallest $q$-value at which $S(q)=0$. Although $q_c$ is not well defined
for finite $N$, it can be estimated reliably. Our algorithm deteriorates only when the network is too 
small (in this case, however, an extensive search of the optimal solution can be performed). 
In general, the smaller $q_c$, the more effective is the corresponding strategy, since the epidemic can be
prevented by vaccinating a smaller set of nodes.

The identification of superspreaders is in general an NP-complete problem~\cite{Kempe_2003}, and most 
likely this is also true for finding superblockers. 
Therefore, heuristic approaches have to be used. Typically a score is assigned to 
each node using local~\cite{Liu_Chaos2016,Holme_EPL2004_LocalVaccination} or 
global~\cite{Holme_PRE2002_AttackVulnerability,Schneider_EPL2012,Morone_Nature2015} properties.
In contrast to most previous papers, we use an ``inverse" \cite{Schneider_EPL2012} strategy.
We start from a configuration where all nodes are considered as potentially ``dangerous'' and are thereby virtually vaccinated ($q=1$);
then, increasingly dangerous nodes are progressively ``unvaccinated''  (i.e. made susceptible). 
This is directly related to the concept of {\it explosive percolation} (EP) proposed by Achlioptas 
{\it et al.}~\cite{Achlioptas_Science2009}, 
\footnote{ At variance with Ref.~\cite{Achlioptas_Science2009}, where {\it bond} percolation has been explored,
here, it is more natural to deal with {\it site} percolation: this is a, however, a  minor difference.}.
EP has been discussed in a large number of papers because of its very unusual threshold 
behavior~\cite{Saberi_PhysRep2015_ReviewPercolation}. It is 
reminiscent of a wide range of ``explosive" (i.e. strongly discontinuous) phenomena in natural processes 
like social contagion~\cite{GomezGardenes_PerezReche_SciRep2016}, 
generalized epidemics~\cite{Janssen_PRE2004,Bizhani_PRE2012,Chung_PRE2014}, 
$k$-core percolation~\cite{dorogovtsev2006k},
interdependent networks~\cite{Buldyrev_Nature2010,Son-Grassberger_EPL2012},
synchronization~\cite{Gardenes:2011,Leyva:2012,Motter:2013,Kurths:2013} or 
jamming~\cite{Echenique2005} but so far no application of EP had been proposed. 
To our knowledge, immunization is the first context where EP is practically used.

Two other ingredients are also essential to make our method fast and efficient: (i) We use two different 
schemes for $q>q_c$ and $q<q_c$, which both combine local and quasi-global information; (ii) We use the 
fast Newman-Ziff algorithm \cite{Newman-Ziff_PRE2001} to identify clusters of susceptible nodes. In addition, we use a number of heuristic tricks that will be described below.

In the following we test the performance of EI for both real-world and model networks. 
Overall, it gives the smallest values of ${S}(q)$ (although other strategies my locally perform better
for specific $q$-values).
Moreover, it gave in all cases by far the lowest values of $q_c$ compared to all other strategies, except for the 
very recent message passing algorithms of \cite{braunstein2016network,mugisha2016identifying}.   
Following the mainstream in network studies,
we focused on ${S}(q)$, which corresponds to outbreaks starting in ${\cal G}(q)$. 
However, outbreaks can also start in any other cluster. An improved success measure $\bar{S}(q)$
can be indirectly defined from the average number of infected sites
$\langle n_{inf}\rangle = N\sum_i S_i^2(q) \equiv N \bar{S}^2(q)$, where $S_i(q)$ denote the sizes of all clusters,
ordered from the largest to the smallest one ($S_1(q)\equiv S(q)$).  If this is used, our algorithm turns out to be 
yet more efficient, and is optimal even when ${S}(q)$ might suggest that it is not (see Appendix~\ref{app:Sbar}). 
In addition, our algorithm is also extremely fast: Its time complexity is linear in $N$ up to logarithms.

{\bf The method}:
We adopt a recursive strategy. Given a configuration with a mixture of vaccinated and susceptible nodes, $m$ candidates are randomly 
chosen among the vaccinated ones and the least dangerous (i.e. the weakest blocker) is unvaccinated 
(we use typically $m \sim 10^3$~\footnote{Notice that $m=2$ was used 
in \cite{Achlioptas_Science2009}, in the context of EP.}).
The selection process is based on a node score quantifying its blocking ability.
The guiding intuition is that harmless nodes should be identified on the basis of the size of the cluster of 
susceptible nodes they would join if unvaccinated (these clusters should be kept small)
and the local effective connectivity which measures
their potential danger if made susceptible. As the relative importance of these two ingredients is significantly different below and 
above the immunization threshold, we use two different scores.
The details of both definitions were obtained by a mix of heuristic arguments and trial-and-error. 
They should not be considered as essential, and indeed very similar results were obtained by different 
ans\"atze within the same spirit, see Appendices~\ref{app:Sigma1} and \ref{app:sigma2}. 

\begin{figure}
\includegraphics[width=6cm]{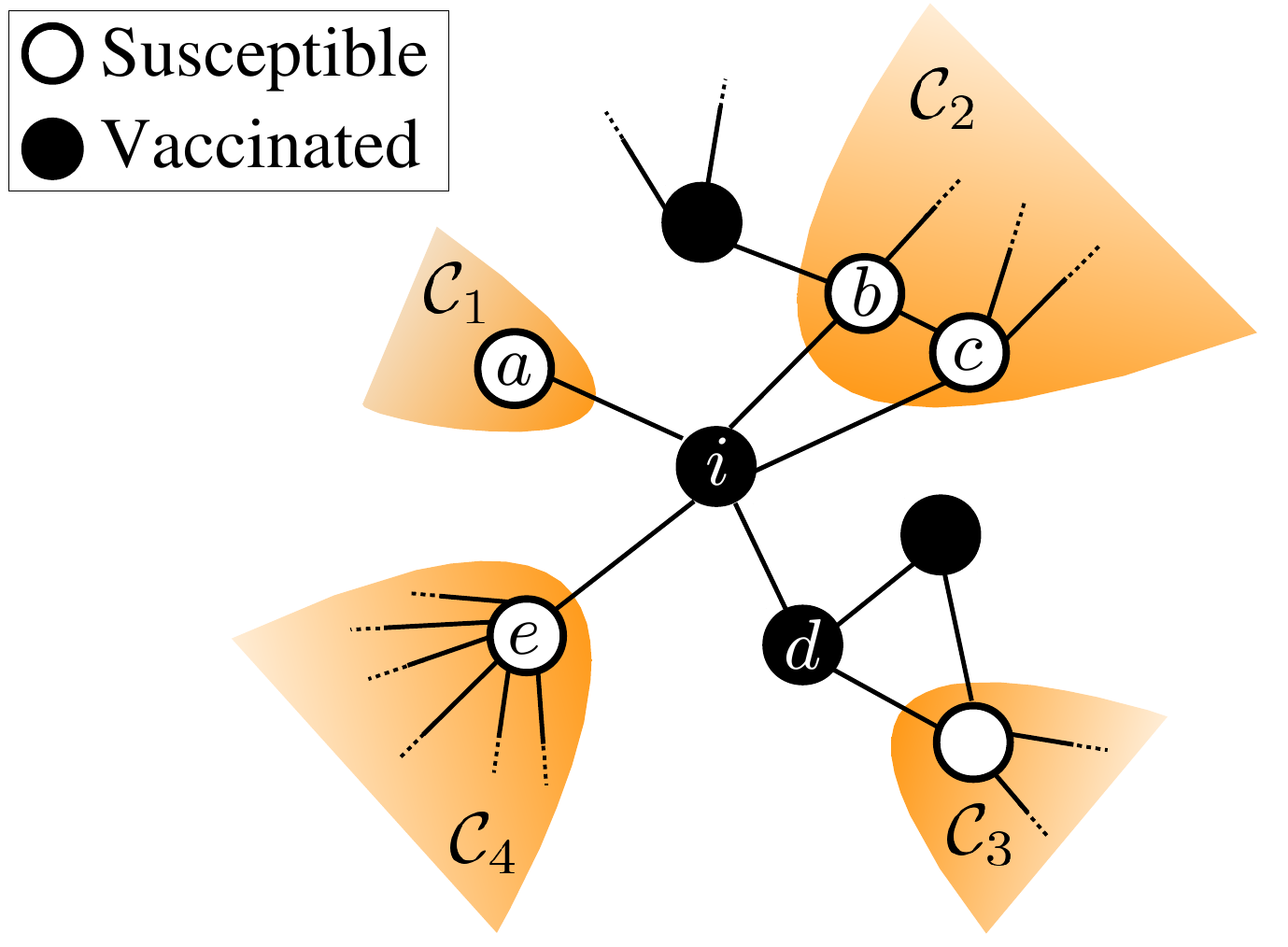}
\caption{\label{Network_Schematic} 
Illustration of the effective degree $k^{(\rm eff)}_i$ of a generic vaccinated node $i$.
Shaded areas identify distinct clusters of susceptible nodes.
With reference to Eq.~(\ref{eq:k_eff}),  $k_i=5$ (see the five neighbors of $i$ labelled
$a-e$); only node $a$ is a leaf, so that $L_i=1$.
Assuming that the cutoff $K$ is set equal to 6, none of the nodes $a-d$ is a hub,
while node $e$ is a hub provided no more than one of its neighbours is a hub itself.}
\end{figure}

The first score, used in the large-$q$ region, is the sum of two separate contributions,
\be
\sigma_i^{(1)} = k_i^{(\rm eff)} + \sum_{{\cal C} \subset {\cal N}_i} (\sqrt{|{\cal C}|}-1)~ .
 \label{eq:score_1}
\ee
The first term $k^{(\rm eff)}_i$ quantifies the potential danger due to the effective local connectivity. 
It is determined self-consistently from the bare degree $k_i$, 
\be
k^{(\rm eff)}_i=k_i-L_i-M_i(\{k^{(\rm eff)}_j\})~.  
\label{eq:k_eff}
\ee
The number $L_i$ of leaves is subtracted since they do not lead anywhere. The number
$M_i$ of {\it strong} hubs is subtracted since, in our inverse strategy, they will likely be vaccinated in
case of an epidemic. The analysis of several networks has led us to identify {\it strong} hubs
in a recursive way as those nodes with $k^{(\rm eff)}_i \geq K$ for a suitably chosen cutoff $K$.
We will see that the best results are typically obtained with $K\approx 6$ for many networks, 
including Erd\"os-R\'enyi (ER) networks within a wide range of $\langle k\rangle$ (see Appendix~\ref{app:Distrib_k}). 
An example of how $k_i^{(\rm eff)}$ is determined is given in Fig.~\ref{Network_Schematic}, where
all of the above details are shown at work.
Notice that, according to Eq.~\eqref{eq:k_eff}, nodes surrounded by hubs may play a minor blocking role 
for spread and can be left unvaccinated, as compared to nodes without hub neighbors. 
This idea  is similar to the score used in \cite{Liu_Chaos2016},
but it is opposite of what is assumed e.g. in page rank \cite{Newman_Book2010} and in the 
collective influence" defined in \cite{Morone_Nature2015}.

The second term on the r.h.s. of Eq.~\eqref{eq:score_1} is a $q$-dependent contribution which takes into 
account the connectivity of the network beyond the neighbours of node $i$.  It is based on the size
of the clusters that would be joined by turning the $i$-th node susceptible:
${\cal N}_i$ is the set of all clusters linked to the $i$th node, while $|{\cal C}|$ is the size of cluster ${\cal C}$.
A question arises about the weight to give to this contribution. In Ref.~\cite{Schneider_EPL2012}, where only the 
nonlocal term was considered, a proportionality to the number $|{\cal C}|$ of nodes was assumed; here we find better 
results by assuming a square root dependence (see also Appendix~\ref{app:Sigma1}). Additionally, our choice preserves a higher fragmentation, 
preventing relatively large clusters of susceptibles to merge together. 
Finally, the nonlocal character of this contribution is better represented by imposing that each addendum is 
larger than zero only for clusters containing strictly more than one node: 
this is the reason for subtracting 1;  numerical simulations confirm the validity of this choice.

As we will see, using $\sigma_i^{(1)}$ yields small values of $q_c$. 
However, it is not suitable to keep a small $S(q)$ below $q_c$. 
This is due to the fact that below $q_c$ it leads to big jumps in $S(q)$ when two large clusters join
(similar jumps were seen in \cite{Schneider_EPL2012,braunstein2016network,mugisha2016identifying}).
As a result of the merging process, many nodes (at the interface between the two clusters)
suddenly become harmless without being treated as such. Accordingly, we use a different 
score $\sigma_i^{(2)}$ with an even stronger opposition to cluster merging,
\begin{equation}
   \label{eq:score_2}
   \sigma_i^{(2)}=
   \begin{cases}
      \infty       &\text{ if } \;\; {\cal G}(q) \not\subset  {\cal N}_i,  \\
      |{\cal N}_i| &\text{ else, if } \arg\min_i |{\cal N}_i| \text{ is unique,}\\
      |{\cal N}_i| + \epsilon |{\cal C}_2| &\text{ else.}
   \end{cases}
\end{equation}
Here $|{\cal N}_i|$ is the number of clusters in the neighborhood of $i$, ${\cal C}_2$ is the second-largest 
cluster in ${\cal N}_i$, and $\epsilon$ is a small positive number (its value is not important 
provided $\epsilon \ll 1/N$). Thus we select only candidates which have 
the giant cluster in their neighborhood; among these we pick the candidate with the smallest number of 
neighboring clusters, and if this is not unique, we pick the candidate for which the second-largest 
neighboring cluster is the smallest (see also Appendix~\ref{app:sigma2}). 
The $q$-value where the performance of $\sigma_i^{(1)}$ deteriorates depends on the network type and 
its size. However, we expect the effect to become more pronounced below a value $q^*$ where 
$S(q^*) \approx 1/\sqrt{N}$, i.e. when a giant cluster starts dominating.

Two remarks are in order about the efficiency of our algorithm:
(i) In \cite{Schneider_EPL2012} all vaccinated nodes were considered as candidates 
to become susceptible during the  de-immunization process. 
This makes the algorithm very slow and prevents its use for large networks. In our tests 
already $m=10$ candidates gave very good results, and using $m=1000$ candidates led to no 
noticeable degradation (see Appendix~\ref{app:NumberCandidates});
(ii) When joining clusters, we used the very fast Newman-Ziff percolation algorithm \cite{Newman-Ziff_PRE2001} 
which has time complexity $O(N)$ for networks with bounded degrees. It also gives, at each 
moment, the size of the largest cluster, whose determination would otherwise 
need most of the CPU time. As a result, we could analyze networks 
with $10^8$ nodes within hours on normal workstations.

\begin{figure}
\begin{centering}
\includegraphics[scale=0.3]{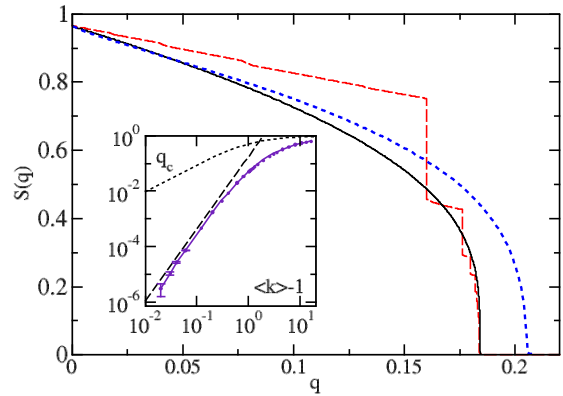}
\par\end{centering}
\caption{\label{ER-3.5.fig} Relative size $S(q)$ of the largest clusters against $q$, for 
  ER networks with $N=10^6$ and $\langle k\rangle = 3.5$. The dashed curve with jumps is 
  obtained, if EI is used with score $\sigma^{(1)}$ for all $q$, 2000 candidates, and $K=6$.
  The continuous curve is obtained with $\sigma^{(2)}$ for $q<q^*$, where $S(q^*) = 1/500$.
  The dotted line shows the result from \cite{Morone_Nature2015}. The inset
  shows a log-log plot of $q_c$ against $\langle k\rangle-1$. The straight line indicates the 
  power law $q_c \sim (\langle k\rangle-1)^{2.6}$, while the dotted curve shows the result
  for random immunization. }
\end{figure}

{\bf Numerical results:} 
As a first test we studied ER networks with average degree 
$\langle k\rangle = 3.5$ (to compare with results from \cite{Morone_Nature2015}). 
Overall, the best results are obtained by using the scores given in 
Eqs.~(\ref{eq:score_1}) and (\ref{eq:score_2}) (Fig.~\ref{ER-3.5.fig}, solid line in main plot).
The dashed line is obtained by using $\sigma^{(1)}_i$ for all $q$ (the big jumps, which were also
seen in \cite{Schneider_EPL2012}, correspond to joinings of big clusters). It is in general 
worse than the continuous curve, except close to the jumps (see, however, Appendices~\ref{app:Sigma1} and \ref{app:sigma2}). 
Finally we show in Fig.~\ref{ER-3.5.fig} also the results obtained with the 
recently proposed collective influence algorithm \cite{Morone_Nature2015}, which was hailed 
in as ``perfect" \cite{Kovacs-Barabasi_Nature2015}. They are significantly worse. Our estimate
$q_c \leq 0.1838(1)$ is also smaller than the best estimate $0.192(9)$ obtained in
\cite{Morone_Nature2015} using extremal optimization~\cite{Boettcher-Percus_PRL2001}, and used 
there as ``gold standard" for small networks (it is too slow to be used for large networks).

As regards ER networks with other values of $\langle k\rangle$, we first looked at 
$\langle k\rangle=4$, since this had been used in \cite{Schneider_EPL2012}. Our results are 
similar to those of \cite{Schneider_EPL2012}, but significantly better. Next we estimated
$q_c$ for a wide range of $\langle k\rangle$. By using networks with $N$ up to $2^{24}$ we were 
able to obtain precise results even for $\langle k\rangle$ very close to the threshold 
$\langle k\rangle=1$ for the existence of a giant cluster. The results, shown in the inset of 
Fig.~\ref{ER-3.5.fig}, suggest that $q_c$ satisfies for small $\langle k\rangle$ a power law
\be
   q_c \sim (\langle k\rangle-1)^{2.6},
\ee
where the error of the exponent is $\approx \pm 0.2$. This should be compared to random
immunization~\cite{Cohen_PRL2001}, $q_c^{\text{rand}}=(\langle k \rangle-1)/\langle k \rangle$
(dotted curve in the inset of Fig.~\ref{ER-3.5.fig}).
The difference in the exponents reflects the fact that a nearly critical cluster can be 
destroyed by removing a few ``hot" nodes, whence targeted attacks become more efficient as 
$\langle k\rangle$ approaches the threshold.

Surprisingly, for all $\langle k\rangle$ values except very close to 1, best results are obtained with
$K=6$. This suggests that most nodes with $k_i^{(\rm eff)}>6$ are vaccinated at $q_c$, independently
of the average degree. This was also verified directly: Although there is no strict relationship 
between effective degree and blocking power (some hubs were not vaccinated at $q_c$, while some 
nodes that were vaccinated are not strong hubs), there is a very strong correlation, stronger 
than between actual degree and blocking power (see Appendix~\ref{app:Distrib_k}). On the other 
hand, very few nodes with small $k_i^{(\rm eff)}$ have to be vaccinated (about 1 per mille of the nodes with 
$k_i^{(\rm eff)}=3$), in contrast to claims in \cite{Morone_Nature2015} that weakly connected nodes are
often important blockers.

\begin{figure}
\begin{centering}
\includegraphics[scale=.35]{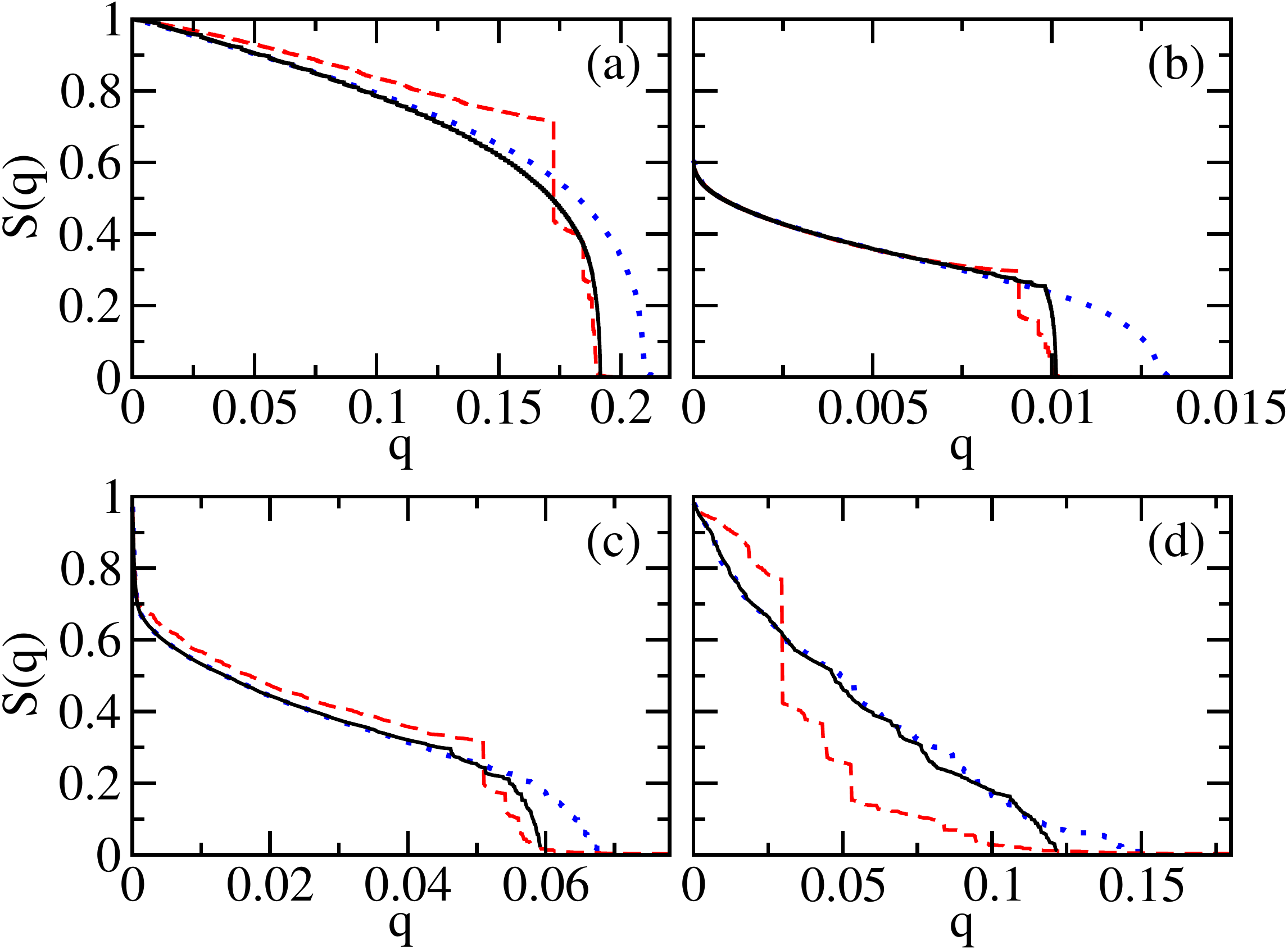}
\par\end{centering}
\caption{\label{ER-BA} Relative size $S(q)$ of the largest clusters in SF networks of size $N=10^6$ obtained with (a) the Albert-Bar\'abasi model ($\gamma=3$) and (b) the configuration model with $\gamma=2.5$. Panels (c) and (d) show results for the cattle and airport transportation networks, respectively. 
Different line types correspond to different algorithms: EI using scores $\sigma_i^{(1)}$ and $\sigma_i^{(2)}$  (continuous line) or only score $\sigma_i^{(1)}$ (dashed line) and the algorithm in \cite{Morone_Nature2015} (dotted line).}
\end{figure}

{\it Scale-free networks}. EI also gives excellent results for scale-free (SF) networks with 
node degree distribution $p_k\sim k^{-\gamma}$, built with both the Barab\'asi-Albert method 
(fixed $\gamma=3$) and the configuration model ($\gamma$ can be 
tuned)~\cite{Barabasi-Albert_Science1999,Newman_Book2010}. Our results are significantly better 
than those obtained with the method from \cite{Morone_Nature2015} for both settings 
(Fig.~\ref{ER-BA}(a) and (b)). Using a single score across the entire $q$-range gives again 
the best estimate for $q_c$, while the two-score strategy proves generally superior for $q<q_c$.
The jumps obtained in the single-score strategy are less pronounced for the configuration model 
(and thus the two-score strategy seems less preferable), but the superiority of the 
two-score strategy becomes again clear when using the improved $\bar{S}(q)$ discussed in Appendix~\ref{app:Sbar}.

Observe that the shape of $S(q)$ near $q=0$ is concave/convex for large/small $\gamma$ 
(compare panels (a) and (b) in Fig.~\ref{ER-BA}). The convex shape for small $\gamma$ is 
due to the presence of many hubs which lead to a drastic decrease of $S(q)$ when vaccinated at small $q$.

{\it Real world networks}: We have also studied the performance of EI on a number of real-world networks, starting 
from an example in which immunization plays an important role for food 
security~\cite{Keeling_Nature2003,Kao_JRoyInterface2007}: a network of Scottish cattle movements \cite{cattle_data}. The network consists of $N=7228$ premises (nodes) connected by $E=24784$ transportation events (edges) occurring between 2005 and 2007. The node distribution obeys a power-law with exponent $\gamma=2.37\pm 0.06$ (Maximum likelihood fit). The scenario is similar to that of SF networks with small $\gamma$ (compare panels (c) and (b) in Fig.~\ref{ER-BA}).
Again, $S(q)$ decreases quite quickly because of the presence of many well connected nodes (e.g. markets and slaughterhouses), 
whose immunization leads to a drastic decrease of the largest cluster.
Once again we see that our strategy using two scores is superior to the previous approaches.

We have also studied several networks that were used as benchmark in previous works. This includes the high-energy 
physicist collaboration network \footnote{See http://vlado.fmf.uni-lj.si/pub/networks/data/hep-th/hep-th.htm} and 
the internet at autonomous system level \footnote{See http://www.netdimes.org}. 
In both cases our results are similar to, but slightly better than, in \cite{Schneider_EPL2012} 
(which were the best previous estimates).
The results for these and soil networks~\cite{PerezReche_PRL2012_Soil} are shown in Appendix~\ref{app:RealNetworks}.

A particularly problematic case is the airline network~\cite{Airports_data}, also studied in \cite{Schneider_PRE2011}. 
This is a rather small network ($N=3151$ and $E=27158$) with a broad degree distribution (power-law with $\gamma=1.70 \pm 0.04$).
The results reported in Fig.~\ref{ER-BA}(d) show that $\sigma_i^{(1)}$ provides very low $S(q)$ almost everywhere.
We conjecture this is due to the unusually small $\gamma$, which implies an abundant number of hubs. 
As a result, the outcome of the score $\sigma_i^{(2)}$ strongly depends on the value of $q_c$ that is selected.
It is anyway clear that a suitable combination of them provides the optimal results.

{\bf Conclusions:} In this paper, we extend the explosive percolation concept to propose a two-score strategy for attacking networks that proves superior to all previously proposed protocols. 
The comparison between the two scores suggests that an everywhere optimal strategy using a single score is unlikely to exist. This is to be traced back to the NP completeness of the problem.
Since immunization of a network by vaccinating nodes can be regarded as a strategy for destroying the 
network on which an infection can propagate, this also gives a nearly optimal strategy for immunization. Our explosive 
immunization method seems superior, both as regards speed and minimal cost (as measured by the number of vaccinated nodes) 
to all previous strategies. 

We have focused on immunization of nodes but EI can also be applied to immunization of links. This would provide nearly optimal quarantine strategies which might significantly improve the typical brute-force implementation which cut all the links between two parts of a network. Targeted removal of links with high betweeness centrality is the basis for one of the most efficient algorithm for finding network communities~\cite{Girvan-Newman_PNAS2002}. We propose that explosive immunization of links should also provide a very efficient algorithms for community detection.

The authors acknowledge financial support from the Leverhulme Trust (Grant No. VP2-2014-043) and from Horizon2020 (Grant No. 642563 - COSMOS).

\appendix
\section{Improved success measure for network immunization}
\label{app:Sbar}
The standard measure for the success of an immunization strategy is the relative size of the largest 
connected cluster, $\mathcal{G}(q)$, after a fraction $q$ of nodes have been vaccinated, 
\be
   S(q) =  |{\cal G}(q)|/N~,   \label{S1}
\ee
where $|{\cal G}(q)|$ is the size of the largest cluster. The motivation for this is that a strongly 
infective disease that hits a random node will in average
infect a region of size $NS(q)^2$ in the largest cluster, where the first factor of $S(q)$ is for
the probability that the largest cluster is hit at all, and the other factors give the number of
infected sites, if it does so.
This quantity neglects the effect of smaller clusters, following a widespread habit in network science. 
Often this is justified because their contribution is small and/or hard to 
estimate. But in the present case, the contribution of clusters other than the largest one can be 
substantial, and it can be taken into account easily. 
In order to incorporate the effect of epidemics starting in all clusters in our analysis, 
let us assume the clusters to be ordered by size, $|{\cal G}(q)| \equiv |{\cal C}_1(q)| \geq |{\cal C}_2(q)| \geq \ldots$. 
The probability that a random outbreak starts in a cluster ${\cal C}_i$ is $S_i(q) = N^{-1} | {\cal C}_i |$ 
and its maximum size is $N S_i(q)$. 
Accordingly, the average number of infected sites in an random outbreak is 
\be
   \langle n_{\rm infected}\rangle = N S(q)^2 + N \sum_{i\geq 2} S_i(q)^2 = N{\bar{S}}(q)^2, 
\ee
where, perturbatively,  
\be
\bar{S}(q) = S(q) \left[ 1+ \frac{1}{2}\sum_{i \geq 2} \left(\frac{S_i(q)}{S(q)}\right)^2 + \ldots \right]
\ee

\begin{figure}
\begin{centering}
\includegraphics[width=8cm]{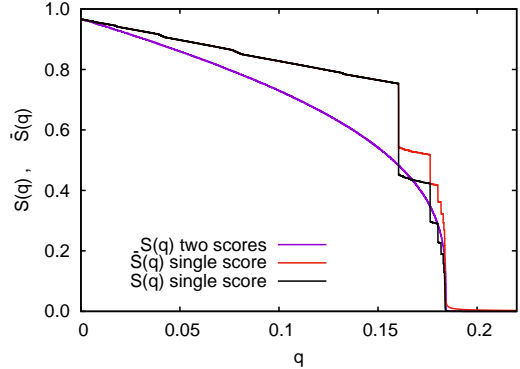}
\par\end{centering}
\caption{\label{figS1}  Success measures $S(q)$ (lower (black) steplike curve) and  $\bar{S}(q)$
  (upper (red) steplike curve) for ER networks with $\langle k\rangle = 3.5$, if score $\sigma^{(1)}$ is 
  used for all $q$. The smooth curve corresponds to the case where $\sigma^{(2)}$ is used
  for $q<q_c$. In that case, the curves for $S(q)$ and $\bar{S}(q)$ are indistinguishable except in 
a small region $q>q_c$.}
\end{figure}

For EI with both scores there will never be more than one large cluster
(at least for large networks with a well defined $q_c$),
since there is no large cluster for $q>q^*$, and for $q<q^*$ the 
growth of a second large cluster is suppressed. In this case $\bar{S}(q)$ is practically the same 
as $S(q)$, as we indeed checked for ER networks. This is not true, however, for $q<q^*$ if the score
$\sigma^{(1)}$ is used also there. In that case there are in general more than one large cluster,
and $\bar{S}(q)$ is considerably larger than $S(q)$, see Fig.~\ref{figS1}. 

Thus even when it seems better to use $\sigma^{(1)}$ for all $q$ (according 
to the success measure $S(q)$), a more refined success measure might show that the strategy 
of using both scores $\sigma^{(1)}$ and $\sigma^{(2)}$ is superior. 

\section{Motivation for score $\sigma^{(1)}$}
\label{app:Sigma1}
In \cite{Achlioptas_Science2009}, two scores were considered for EP: the `sum rule' where the score is equal to the sum
of the masses of the two joined clusters (remember that in EP links are added, not nodes), and the 'product 
rule', where the score is their product. Alternatively, we could implement the product rule as 
the sum of the logarithms of the cluster masses. In EI we would just replace the two clusters by the set of
all clusters joined by adding the chosen candidate node. When testing these on Erd\"os-R\'enyi (ER) networks 
with $\langle k \rangle= 1.75$, both gave results for $q_c$ that were about $\approx$ 5 to 10\% worse that our 
best result shown in Fig.~2 of the main text, but comparable with the result of \cite{Morone_Nature2015}.

\begin{figure}
\centerline{
\includegraphics[width=9cm,clip]{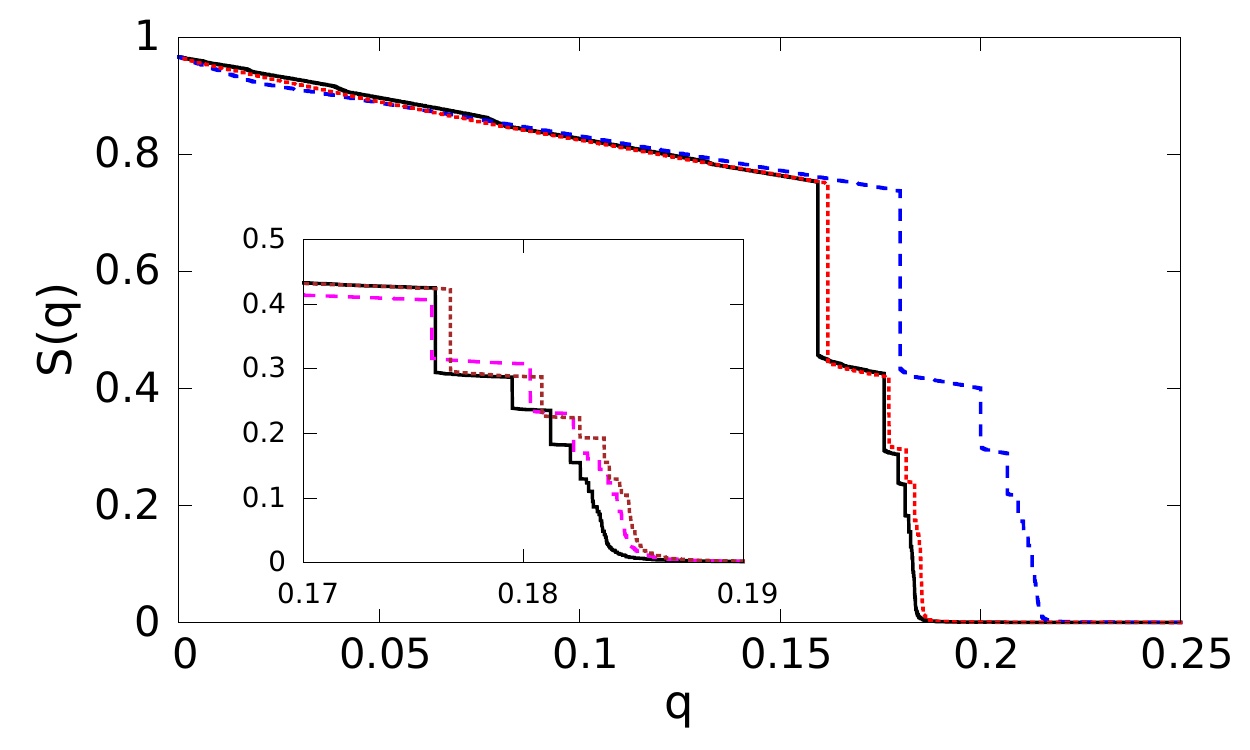}}
\caption{\label{fig:Score1} Comparison of several different choices for score $\sigma^{(1)}$. The main panel shows the relative size $S(q)$ of the largest clusters against $q$ for ER networks with $N=10^5$ and $\langle k\rangle = 3.5$ using $m=2000$ candidates. The continuous black line shows $\sigma^{(1)}$ with $K=6$. The red dotted line is a modified version of Eq.(1) in the main text without the square root. The blue dashed line shows the effect of removing the contribution $k^{(\rm eff)}$ in $\sigma^{(1)}$. The inset shows the effect on $S(q)$ of using Eq.~(1) with different values of the hub cut-off $K$ in the same graph as in the main panel ($K=6,8,10$ for black continuous, magenta dashed and brown dotted lines, respectively).}
\end{figure}

To improve on this, we considered the following heuristics:
\begin{itemize}
\item While the sum rule puts too much weight on very large clusters that are joined, the product rule 
does not put enough weight on them. As a compromise, we added the square roots of the masses, which improved
already slightly the estimate of $q_c$ (compare the continuous and dotted lines in Fig.~\ref{fig:Score1}).
\item Neither of the three above rules pays enough attention to the degree of the chosen node $i$. Thus we 
added a term $\propto k_i$. When trying different relative weights, we obtained best results with the weights
given in Eq.~(1) of the main text. The improvement was again small (1-2\%), but significant.
\item When calculating the degree of the chosen node $i$, neighboring leaves should be dismissed as they would 
not help any epidemic spreading. This was verified, and again the effect was small.
\item If one of the neighbors is a very strong hub, it will finally presumably be vaccinated, in which case 
it cannot contribute to epidemic spreading. It, thereforei, should not be included in the effective neighbor 
count and, maybe, it should not be included in the sum over the neighboring cluster masses (the second term 
in Eq.~(1) of the main text) either. We found (again for ER networks with $\langle k \rangle= 3.5$) that 
the second option had practically no effect (it was neither beneficial nor detrimental), but excluding such
hubs from the effective degree gave a significant improvement, provided the hubs were properly identified 
(compare the continuous and dashed line in Fig.~\ref{fig:Score1}).
The latter included that we define the effective node degree recursively, and the recursion converged
only then in all cases, if it was done with forward substitution. 
The effect of the cut-off $K$ to define hubs is illustrated in the inset of Fig.~\ref{fig:Score1}.
\item In \cite{Schneider_EPL2012}, where a strategy very similar to ours was proposed, the scores were computed by 
dismissing all nodes from the network that were already judged as harmless. We found this to be 
detrimental in most cases, whence we did not use it.
\end{itemize}

\section{Motivation for score $\sigma^{(2)}$}
\label{app:sigma2}
The main motivation for not using $\sigma^{(1)}$ for $q<q_c$, i.e. when there exists already a macroscopic
giant cluster of non-vaccinated nodes, is the occurrence of big jumps in $S(q)$. They occur when two large 
clusters finally have to join. The same jumps were seen also in \cite{Schneider_EPL2012} and in \cite{braunstein2016network,mugisha2016identifying}.
While $S(q)$ is very small for $q$ immediately before the jump, it is much larger than in an optimal 
strategy after the jump. In general, thus, score $\sigma^{(1)}$ gives optimal results for {\it some}
regions of $q$, but these regions are very small and outweighed by the much larger regions where it 
is definitely suboptimal. This conclusion is even strengthened when using the more realistic $\bar{S}$
as a success criterion.

Otherwise said, after two large clusters have joined and the relative size of the largest cluster is $S$,
the fraction $q$ of nodes not yet declared as harmless is much larger (when using $\sigma^{(1)}$) than
necessary. This is fairly easy to understand. As long as the two clusters are still disjoint, all nodes
in the interface between them are `dangerous' in the sense that infecting any of them would give rise to
a large increase of infected area. But after the two clusters have joined, all these nodes are harmless,
even if they are not yet declared as such.
At the same time, these nodes will in general not be connected with each other, and when vaccinating 
them we would effectively influence a single node -- which is of course extremely inefficient.

Thus the guiding principle for $q<q_c$ should be that we want to avoid as much as possible the formation
of isolated, not yet vaccinated, nodes. Only as a secondary criterion we want to avoid the formation of 
large clusters. Equation (3) of the main text does exactly that. 
  
We should add that score $\sigma^{(2)}$ has a very similar effect as the procedure used by 
Morone {\it et al.} \cite{Morone_Nature2015} in their `second pass'. Remember that Morone {\it et al.} first used the 
`collective influence' in a forward strategy, where they identified nodes to be vaccinated. They stopped
this forward iteration when $q\approx q_c$, and then added a second pass (explained only in their supplementary
material) where they `de-vaccinated' some nodes. It was indeed largely this second step which made their
algorithm successful, but they gave no argument for the specific algorithm used for the de-vaccination.

\section{Degree distributions of vaccinated nodes at $q=q_c$}
\label{app:Distrib_k}
\begin{figure}
\includegraphics[scale=0.245,clip]{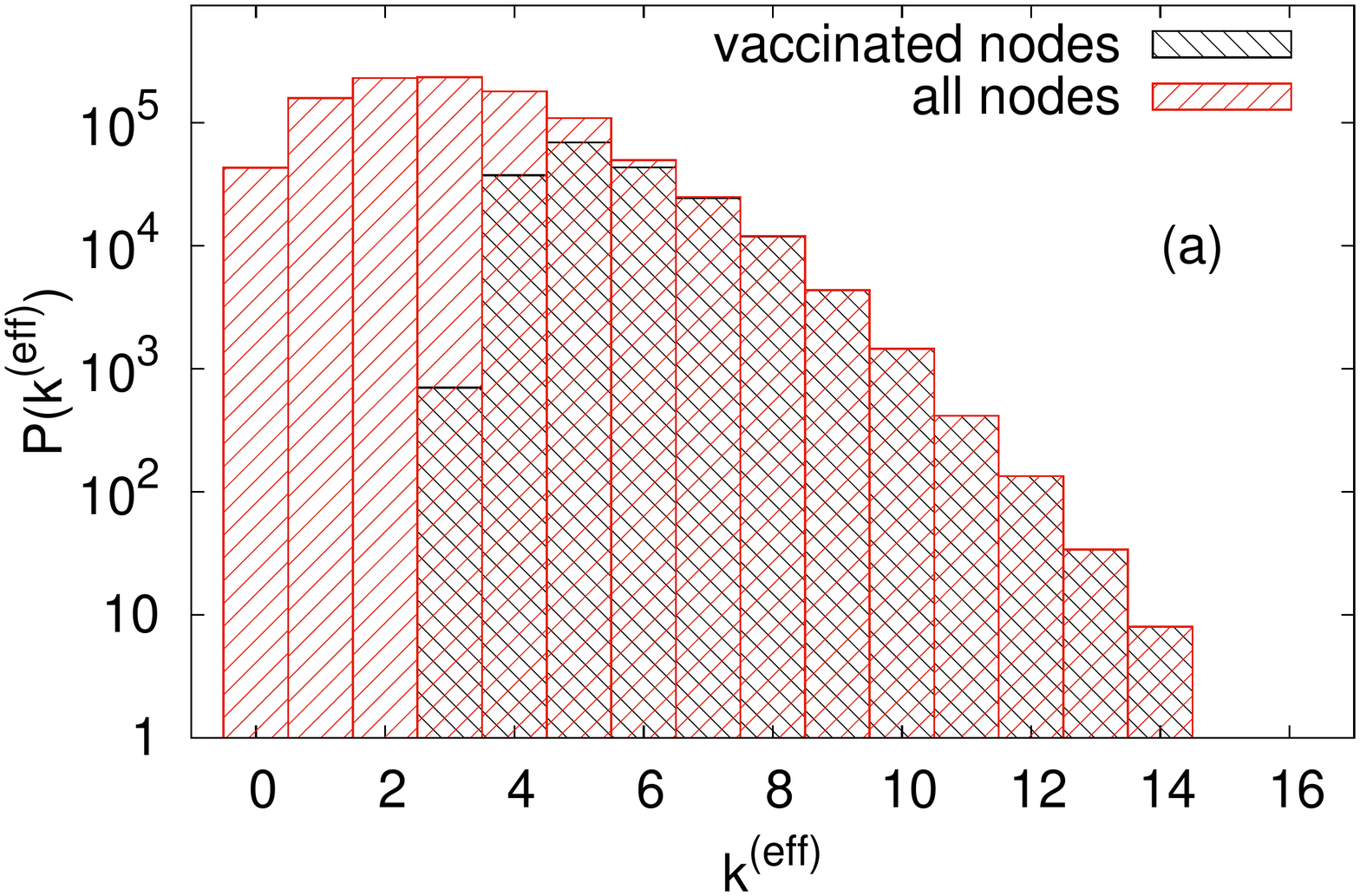}\\
\vskip-20pt
\includegraphics[scale=0.245,clip]{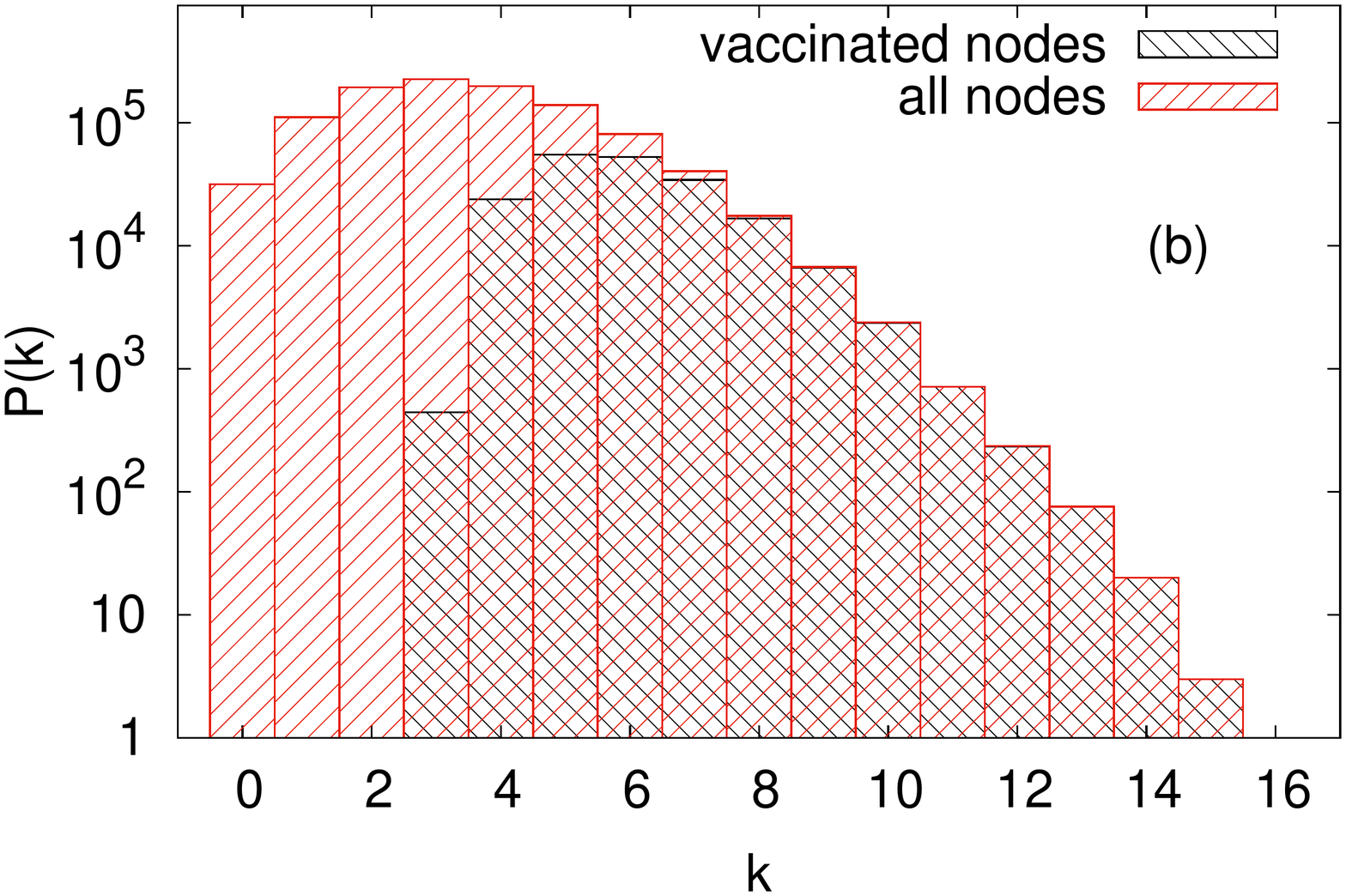}
\vskip-20pt
\caption{\label{figS2} (a) Log-linear plot of $P(k^{({\rm eff})})$ for ER networks
  with $\langle k\rangle = 3.5$. The left histogram is for all nodes, the right one is for those nodes
  that are not declared as ``harmless" at $q=q_c$ and which therefore must be vaccinated in order to
  immunize the network. Panel (b) shows the analogous distributions for the actual degrees. Notice that
  for the vaccinated nodes, the distribution of $k^{({\rm eff})}$ has a slightly sharper cut-off than that 
  of $k$, indicating that $k^{({\rm eff})}$ is a better indicator for nodes that must be vaccinated
  than $k$. The same was found also for all other values of $\langle k\rangle$.}
\end{figure}

Naively, one expects that it is the strongest hubs that should be vaccinated first, but the fact that
network immunization is non-trivial shows that this is not exactly true. If the motivation that led 
us to define the 
effective degree $k^{({\rm eff})}$ is correct, we should expect the vaccinated nodes to be more strongly
concentrated in the high-$k^{({\rm eff})}$ region, than they are concentrated in the high-$k$ region.
Here we show data that indeed confirm this, although the difference is rather small. More precisely,
we show in the left panel of Fig.~\ref{figS2} two histograms: The $k^{({\rm eff})}$-distribution of 
{\it all} nodes in an ER network with $\langle k \rangle = 3.5$ and the distribution of those nodes
that are vaccinated at $q=q_c$. In the right panel the corresponding two $k$-distributions are shown.

We see that in both cases nearly all nodes with degree $>7$ are vaccinated, while nearly all nodes 
with degree $<4$ are left unvaccinated. This agrees with our findings that $K=6$ is optimal in this 
case. A closer look shows that the $k^{({\rm eff})}$-distribution of vaccinated nodes has indeed a
slightly sharper cut off than the $k$-distribution. For instance, while $\approx 25$ \%
of nodes with $k=6$ are not vaccinated, this is true for only $\approx 10$ \% of nodes with
$k^{({\rm eff})}=6$.

\section{Dependence on the number of candidates, $m$}
\label{app:NumberCandidates}
\begin{figure}
\subfloat[]{\label{figS_m_score1}\includegraphics[width=6cm,clip]{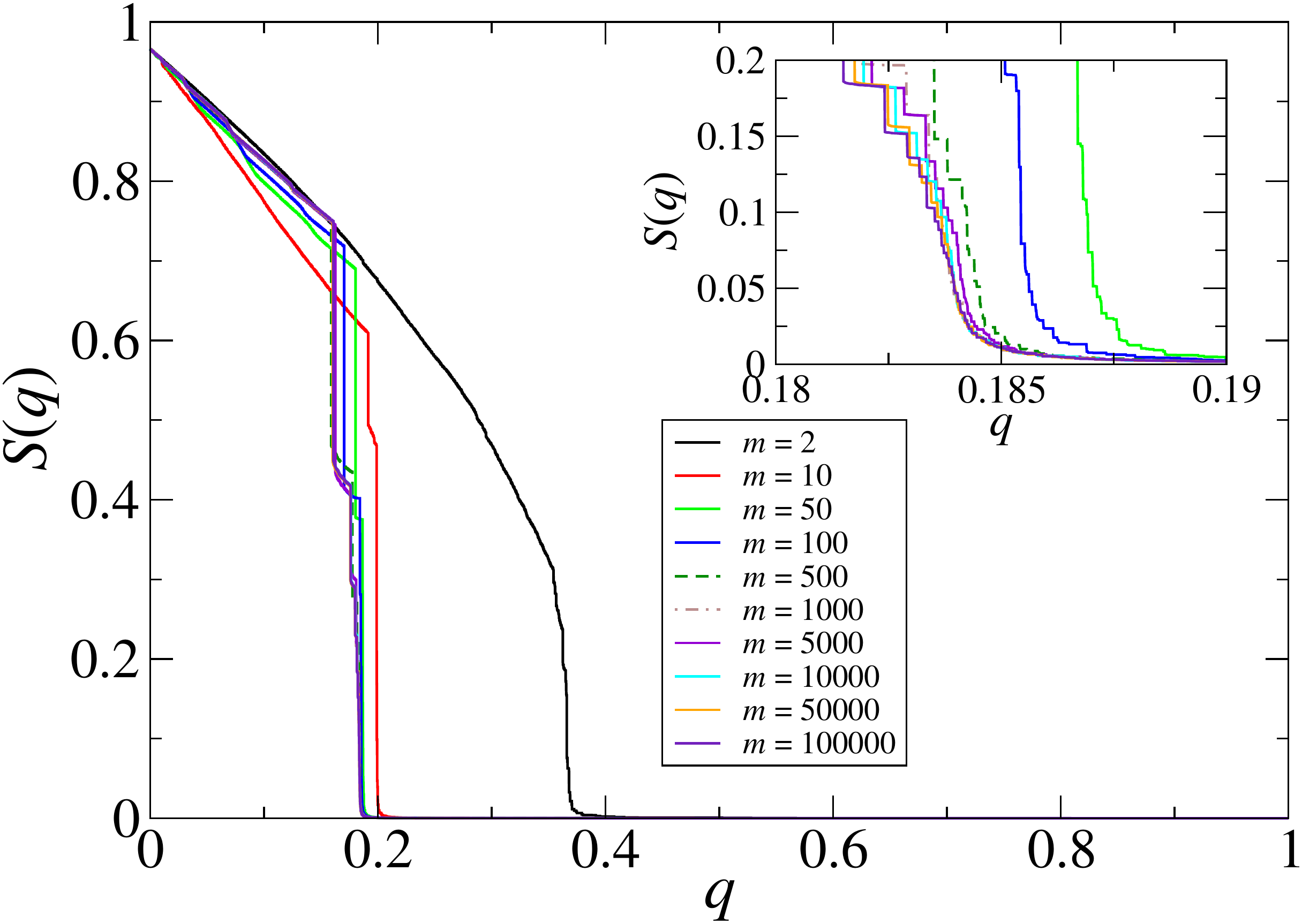}}\\
\subfloat[]{\label{figS_m_score2}\includegraphics[width=6cm,clip]{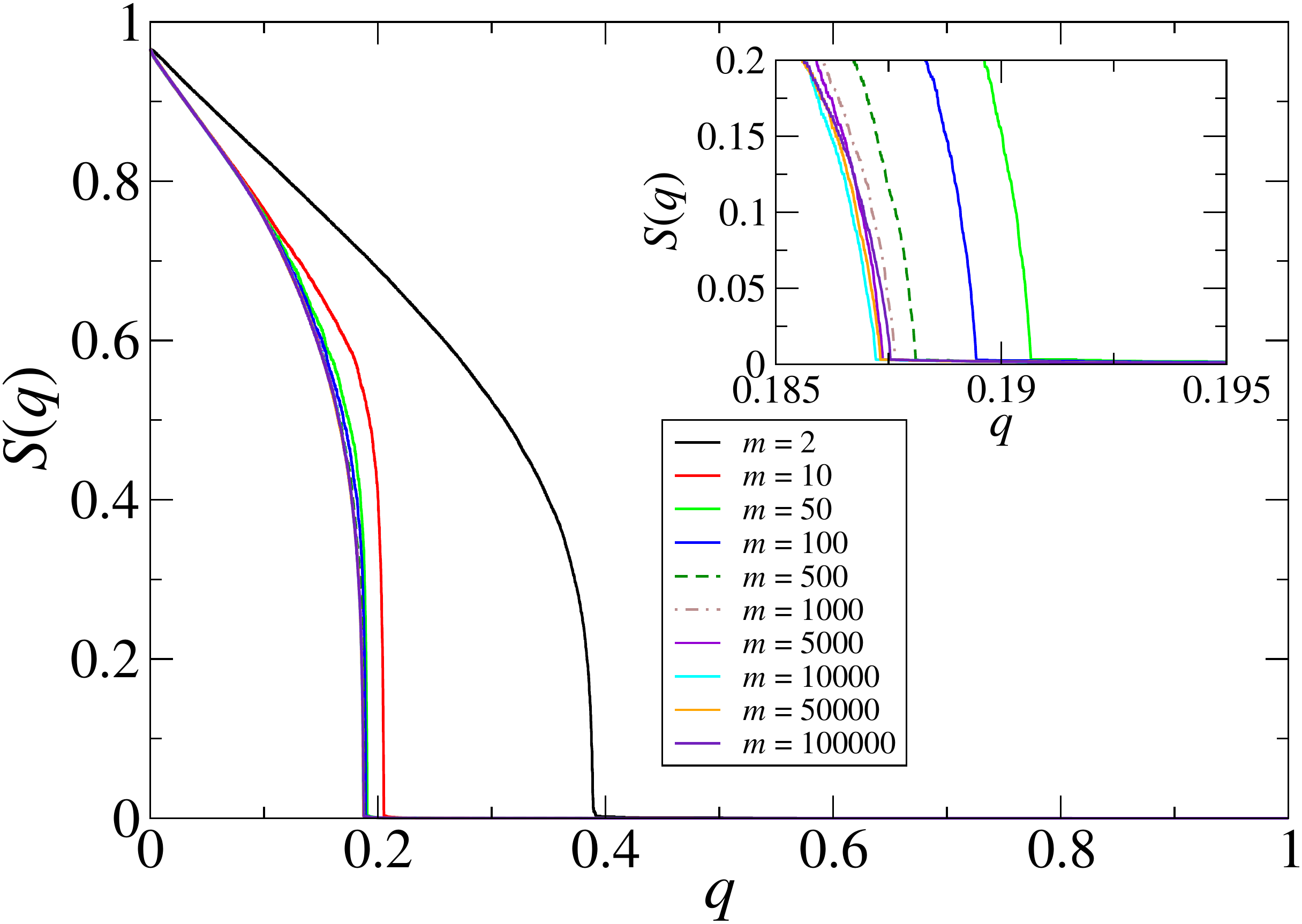}}
\caption{\label{figS_m}  Dependence of $S(q)$ on the number of candidates $m$ for an ER graph with $\langle k \rangle=3.5$ and $N=10^5$. Results in (a) were obtained using score $\sigma^{(1)}$ for all $q$; those in panel (b) were obtained using score $\sigma^{(1)}$ for $q>q^*$ and $\sigma^{(2)}$ for $q<q^*$, where $q^*$ was defined as the point where $S(q)=1/\sqrt{N}$.  Different line types correspond to different values of $m$, as marked by the legend. The inset shows a magnified view of the region around $q_c$. All the curves have been obtained using score $\sigma^{(1)}$ for all values of $q$.}
\end{figure}

\begin{figure}
\centerline{
\includegraphics[width=9cm,clip]{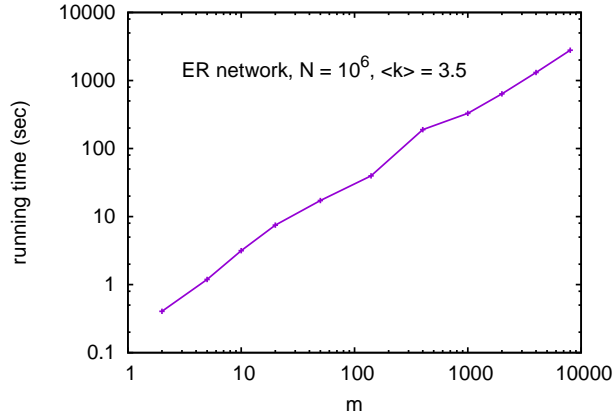}
}
\caption{\label{figS_time_m} Dependence of the running time of the algorithm on the number of candidates $m$ for an ER graph with $\langle k \rangle=3.5$ and $N=10^6$.}
\end{figure}

The number $m$ of vaccinated nodes considered to become susceptible at each step of the de-immunization procedure is a tunable parameter of our model. Fig.~\ref{figS_m} shows the effect of $m$ on the size of the largest cluster, $S(q)$. As can be seen, irrespective of the strategy used to choose the score as a function of $q$, the results become insensitive to the number of candidates already for $m=1000$. The running time increases fast with $m$ (see Fig.~\ref{figS_time_m}). Therefore, using $m \simeq 1000$ represents an important improvement to the running time compared to Ref.~\cite{Schneider_EPL2012} which used $m=N$.

\section{Real-world networks}
\label{app:RealNetworks}
We present the detailed results for three other real-world networks.
In each of them  we compare the two different scores of Explosive Immunization (EI) with
the Collective Influence (CI) method proposed by Morone \emph{et. al.} \cite{Morone_Nature2015}.
We also show an example of how the hub cut-off parameter $K$ in the computation
of the effective degree $k_i^{({\rm eff})}$ modifies the results. In all three plots we use only 
the success measure $S(q)$ (for more easy comparison with previous literature), but we should 
keep in mind that methods producing large steps would become worse when using ${\bar S}(q)$.\\

\begin{figure}
\subfloat[]{\includegraphics[width=8cm]{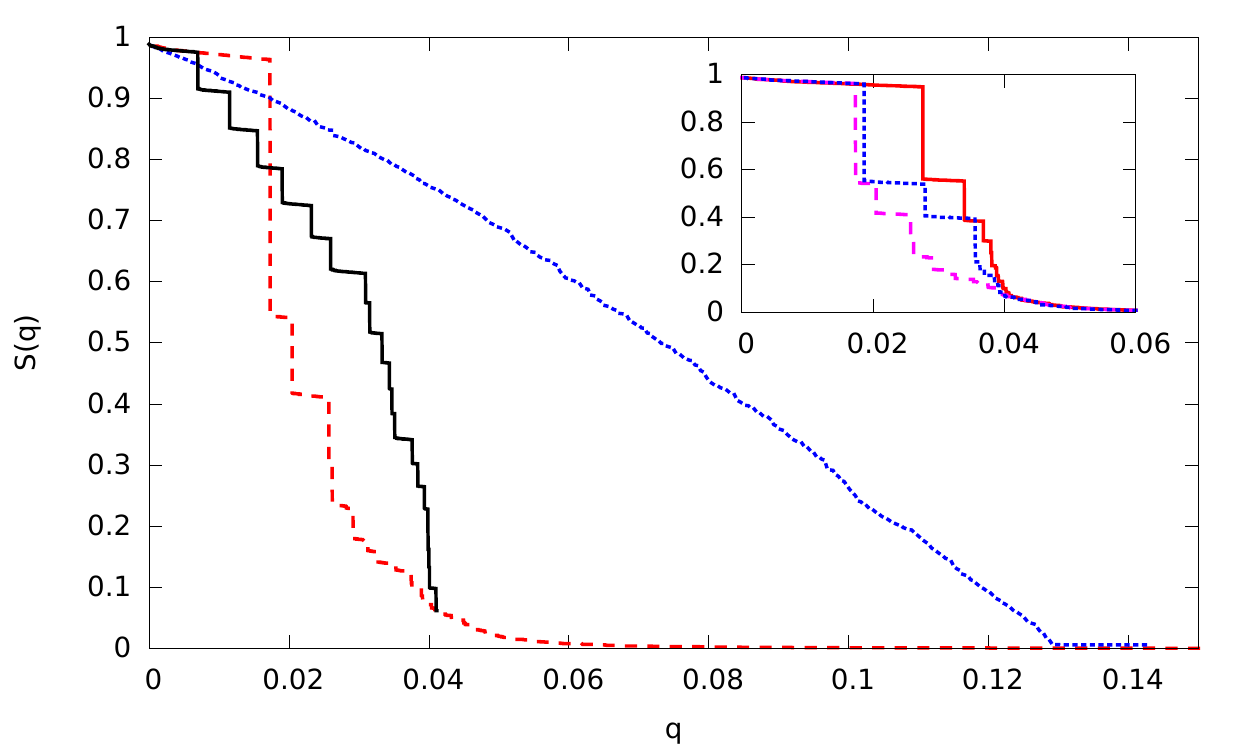}}\\
\subfloat[]{\includegraphics[width=8cm]{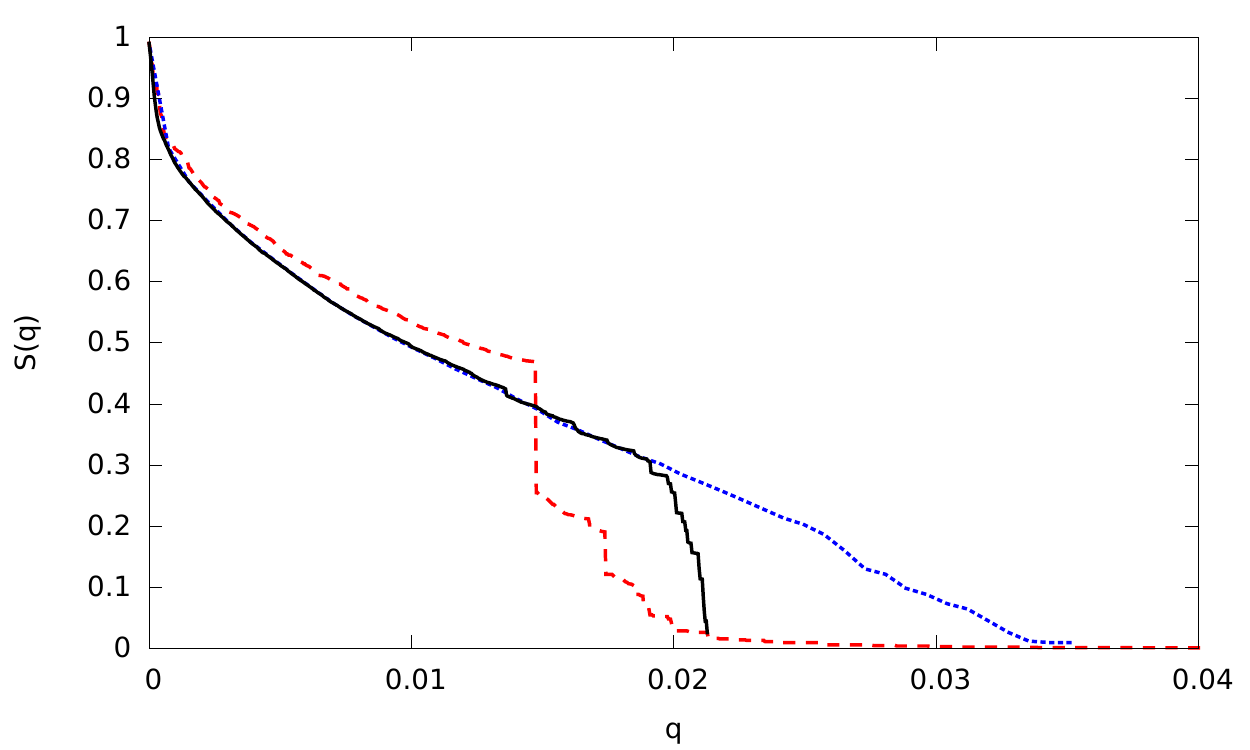}}\\
\subfloat[]{\includegraphics[width=8cm]{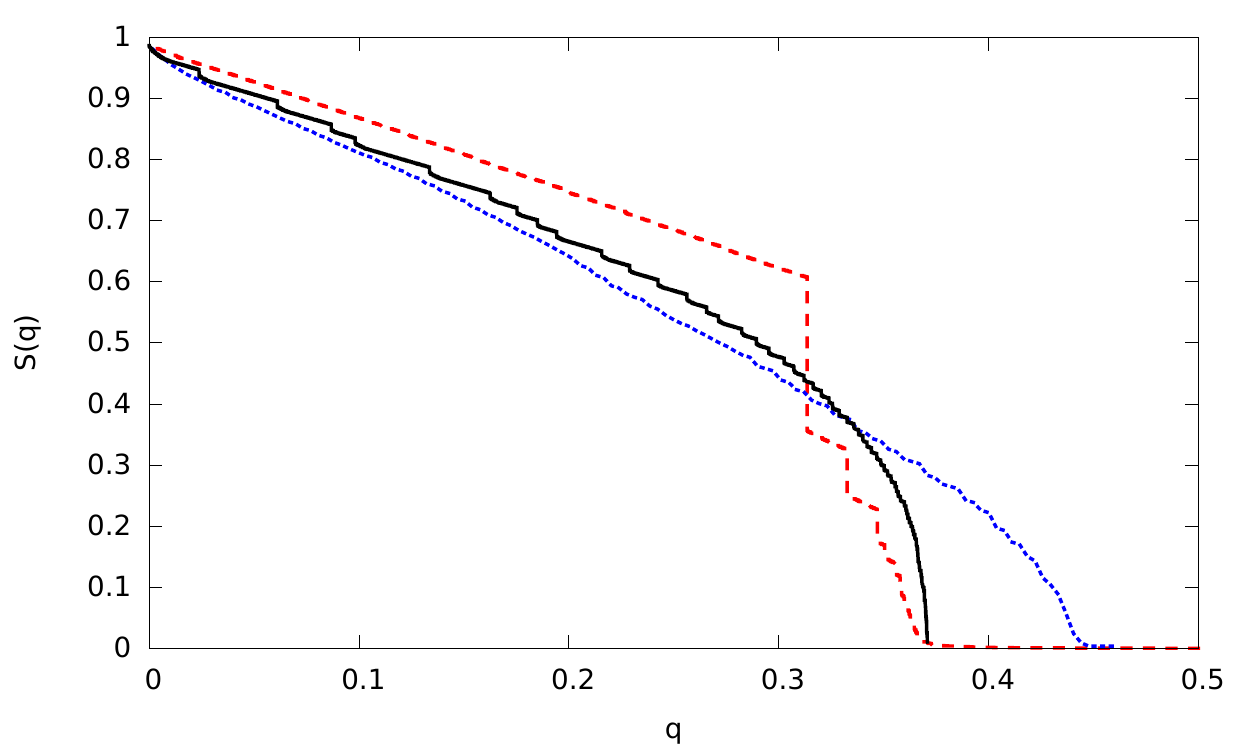}}
\caption{Results of the different real-world networks. The red dashed, black solid and
blue dotted lines corresponds to the $\sigma_i^{(1)}$ score, the $\sigma_i^{(1)}$ \& $\sigma_i^{(2)}$ 
scores, and to the CI method, respectively. Panels (a), (b), and (c) correspond 
to the soil, internet, and high-energy physicist networks discussed in the text.
In the small inset of panel (a) we show different results of $\sigma_i^{(1)}$ 
corresponding to $K=6$ (red solid), 8 (magenta dashed) and 10 (blue dotted).}
\label{fig_real}
\end{figure}

\textbf{Soil network:} We study a network of the structure of soil pores with  $N=49709$ nodes and $E=69563$ links presented 
in \cite{PerezReche_PRL2012_Soil}.
This network has a large clustering coefficient and a limited degree distribution with $\langle k\rangle=2.8$.
In figure \ref{fig_real}(a) we plot the results of EI and CI using $K=8$ and $S(q^*)=0.06$.
In this case, EI produces better results than CI everywhere. 
In particular, using the score
$\sigma_i^{(1)}$ for all $q$ is optimal except when $q$ is very small (see, however, the above 
caveat about using ${\bar S}(q)$ instead of $S(q)$).
In the inset panel we show different values of $K$ effect the outcome of $\sigma_i^{(1)}$ in this network.\\

\textbf{Internet:} In figure \ref{fig_real}b we show the results for a network representing the
Internet at the level of autonomous system\footnote{http://www.netdimes.org} with $N=25612$ and $E=82053$.
We set the parameters $K=6$ and $S(q^*)=0.02$. In this case, different values of 
$K$ do not change significantly the outcome.
In general we observe a behavior similar to the cattle network in which an early vaccination of nodes 
produces a strong decrease of $S(q)$.
Again the EI method gives better results than CI.\\

\textbf{High-Energy Physicists:} Finally, in figure \ref{fig_real}c we use the high-energy physicist 
collaboration network\footnote{ http://vlado.fmf.uni-lj.si/pub/networks/data/hep-th/hep-th.htm}
also used in \cite{Schneider_EPL2012} consisting of $N=27240$ nodes and $E=341923$ links.
We plot the results using $K=6$ and $S(q^*)=0.01$.
The proposed EI algorithm achieves a better value of $q_c$ 
than the one obtained by Schneider \emph{et. al.}. Both are better than the one obtained with CI.
When  the giant component is grown for small $q$ (significantly $ <q_c$), the CI method is similar 
but slightly better than EI. This is the only case that we have found where EI is not optimal everywhere.

\begin{thebibliography}{49}%
\makeatletter
\providecommand \@ifxundefined [1]{%
 \@ifx{#1\undefined}
}%
\providecommand \@ifnum [1]{%
 \ifnum #1\expandafter \@firstoftwo
 \else \expandafter \@secondoftwo
 \fi
}%
\providecommand \@ifx [1]{%
 \ifx #1\expandafter \@firstoftwo
 \else \expandafter \@secondoftwo
 \fi
}%
\providecommand \natexlab [1]{#1}%
\providecommand \enquote  [1]{``#1''}%
\providecommand \bibnamefont  [1]{#1}%
\providecommand \bibfnamefont [1]{#1}%
\providecommand \citenamefont [1]{#1}%
\providecommand \href@noop [0]{\@secondoftwo}%
\providecommand \href [0]{\begingroup \@sanitize@url \@href}%
\providecommand \@href[1]{\@@startlink{#1}\@@href}%
\providecommand \@@href[1]{\endgroup#1\@@endlink}%
\providecommand \@sanitize@url [0]{\catcode `\\12\catcode `\$12\catcode
  `\&12\catcode `\#12\catcode `\^12\catcode `\_12\catcode `\%12\relax}%
\providecommand \@@startlink[1]{}%
\providecommand \@@endlink[0]{}%
\providecommand \url  [0]{\begingroup\@sanitize@url \@url }%
\providecommand \@url [1]{\endgroup\@href {#1}{\urlprefix }}%
\providecommand \urlprefix  [0]{URL }%
\providecommand \Eprint [0]{\href }%
\providecommand \doibase [0]{http://dx.doi.org/}%
\providecommand \selectlanguage [0]{\@gobble}%
\providecommand \bibinfo  [0]{\@secondoftwo}%
\providecommand \bibfield  [0]{\@secondoftwo}%
\providecommand \translation [1]{[#1]}%
\providecommand \BibitemOpen [0]{}%
\providecommand \bibitemStop [0]{}%
\providecommand \bibitemNoStop [0]{.\EOS\space}%
\providecommand \EOS [0]{\spacefactor3000\relax}%
\providecommand \BibitemShut  [1]{\csname bibitem#1\endcsname}%
\let\auto@bib@innerbib\@empty
\bibitem [{\citenamefont {Newman}(2010)}]{Newman_Book2010}%
  \BibitemOpen
  \bibfield  {author} {\bibinfo {author} {\bibfnamefont {M.~E.~J.}\
  \bibnamefont {Newman}},\ }\href
  {http://books.google.co.uk/books?id=q7HVtpYVfC0C} {\emph {\bibinfo {title}
  {{Networks: an introduction}}}}\ (\bibinfo  {publisher} {Oxford University
  Press},\ \bibinfo {address} {Oxford},\ \bibinfo {year} {2010})\BibitemShut
  {NoStop}%
\bibitem [{\citenamefont {Schneider}\ \emph {et~al.}(2011)\citenamefont
  {Schneider}, \citenamefont {Mihaljev}, \citenamefont {Havlin},\ and\
  \citenamefont {Herrmann}}]{Schneider_PRE2011}%
  \BibitemOpen
  \bibfield  {author} {\bibinfo {author} {\bibfnamefont {C.~M.}\ \bibnamefont
  {Schneider}}, \bibinfo {author} {\bibfnamefont {T.}~\bibnamefont {Mihaljev}},
  \bibinfo {author} {\bibfnamefont {S.}~\bibnamefont {Havlin}}, \ and\ \bibinfo
  {author} {\bibfnamefont {H.~J.}\ \bibnamefont {Herrmann}},\ }\href {\doibase
  10.1103/PhysRevE.84.061911} {\bibfield  {journal} {\bibinfo  {journal} {Phys.
  Rev. E}\ }\textbf {\bibinfo {volume} {84}},\ \bibinfo {pages} {061911}
  (\bibinfo {year} {2011})}\BibitemShut {NoStop}%
\bibitem [{\citenamefont {Schneider}\ \emph {et~al.}(2013)\citenamefont
  {Schneider}, \citenamefont {Yazdani}, \citenamefont {Ara{\'{u}}jo},
  \citenamefont {Havlin},\ and\ \citenamefont
  {Herrmann}}]{Schneider_SciRep2013}%
  \BibitemOpen
  \bibfield  {author} {\bibinfo {author} {\bibfnamefont {C.~M.}\ \bibnamefont
  {Schneider}}, \bibinfo {author} {\bibfnamefont {N.}~\bibnamefont {Yazdani}},
  \bibinfo {author} {\bibfnamefont {N.~A.~M.}\ \bibnamefont {Ara{\'{u}}jo}},
  \bibinfo {author} {\bibfnamefont {S.}~\bibnamefont {Havlin}}, \ and\ \bibinfo
  {author} {\bibfnamefont {H.~J.}\ \bibnamefont {Herrmann}},\ }\href {\doibase
  10.1038/srep01969} {\bibfield  {journal} {\bibinfo  {journal} {Sci. Rep.}\
  }\textbf {\bibinfo {volume} {3}},\ \bibinfo {pages} {1969} (\bibinfo {year}
  {2013})}\BibitemShut {NoStop}%
\bibitem [{\citenamefont {Zeng}\ and\ \citenamefont
  {Liu}(2012)}]{Zeng-Liu_PRE2012}%
  \BibitemOpen
  \bibfield  {author} {\bibinfo {author} {\bibfnamefont {A.}~\bibnamefont
  {Zeng}}\ and\ \bibinfo {author} {\bibfnamefont {W.}~\bibnamefont {Liu}},\
  }\href {\doibase 10.1103/PhysRevE.85.066130} {\bibfield  {journal} {\bibinfo
  {journal} {Phys. Rev. E}\ }\textbf {\bibinfo {volume} {85}},\ \bibinfo
  {pages} {066130} (\bibinfo {year} {2012})}\BibitemShut {NoStop}%
\bibitem [{\citenamefont {P{\'{e}}rez-Reche}\ \emph {et~al.}(2010)\citenamefont
  {P{\'{e}}rez-Reche}, \citenamefont {Taraskin}, \citenamefont {Costa},
  \citenamefont {Neri},\ and\ \citenamefont
  {Gilligan}}]{PerezReche_JRSInterface2010}%
  \BibitemOpen
  \bibfield  {author} {\bibinfo {author} {\bibfnamefont {F.~J.}\ \bibnamefont
  {P{\'{e}}rez-Reche}}, \bibinfo {author} {\bibfnamefont {S.~N.}\ \bibnamefont
  {Taraskin}}, \bibinfo {author} {\bibfnamefont {L.~d.~F.}\ \bibnamefont
  {Costa}}, \bibinfo {author} {\bibfnamefont {F.~M.}\ \bibnamefont {Neri}}, \
  and\ \bibinfo {author} {\bibfnamefont {C.~A.}\ \bibnamefont {Gilligan}},\
  }\href@noop {} {\bibfield  {journal} {\bibinfo  {journal} {J. R. Soc.
  Interface}\ }\textbf {\bibinfo {volume} {7}},\ \bibinfo {pages} {1083}
  (\bibinfo {year} {2010})}\BibitemShut {NoStop}%
\bibitem [{\citenamefont {Neri}\ \emph
  {et~al.}(2011{\natexlab{a}})\citenamefont {Neri}, \citenamefont
  {P{\'{e}}rez-Reche}, \citenamefont {Taraskin},\ and\ \citenamefont
  {Gilligan}}]{Neri_JRSInterface2010}%
  \BibitemOpen
  \bibfield  {author} {\bibinfo {author} {\bibfnamefont {F.~M.}\ \bibnamefont
  {Neri}}, \bibinfo {author} {\bibfnamefont {F.~J.}\ \bibnamefont
  {P{\'{e}}rez-Reche}}, \bibinfo {author} {\bibfnamefont {S.~N.}\ \bibnamefont
  {Taraskin}}, \ and\ \bibinfo {author} {\bibfnamefont {C.~a.}\ \bibnamefont
  {Gilligan}},\ }\href {\doibase 10.1098/rsif.2010.0325} {\bibfield  {journal}
  {\bibinfo  {journal} {J. R. Soc. Interface}\ }\textbf {\bibinfo {volume}
  {8}},\ \bibinfo {pages} {201} (\bibinfo {year}
  {2011}{\natexlab{a}})}\BibitemShut {NoStop}%
\bibitem [{\citenamefont {Neri}\ \emph
  {et~al.}(2011{\natexlab{b}})\citenamefont {Neri}, \citenamefont {Bates},
  \citenamefont {Fuchtbauer}, \citenamefont {P{\'{e}}rez-Reche}, \citenamefont
  {Taraskin}, \citenamefont {Otten}, \citenamefont {Bailey},\ and\
  \citenamefont {Gilligan}}]{Neri_PLoSCBio2011}%
  \BibitemOpen
  \bibfield  {author} {\bibinfo {author} {\bibfnamefont {F.~M.}\ \bibnamefont
  {Neri}}, \bibinfo {author} {\bibfnamefont {A.}~\bibnamefont {Bates}},
  \bibinfo {author} {\bibfnamefont {W.~S.}\ \bibnamefont {Fuchtbauer}},
  \bibinfo {author} {\bibfnamefont {F.~J.}\ \bibnamefont {P{\'{e}}rez-Reche}},
  \bibinfo {author} {\bibfnamefont {S.~N.}\ \bibnamefont {Taraskin}}, \bibinfo
  {author} {\bibfnamefont {W.}~\bibnamefont {Otten}}, \bibinfo {author}
  {\bibfnamefont {D.~J.}\ \bibnamefont {Bailey}}, \ and\ \bibinfo {author}
  {\bibfnamefont {C.~A.}\ \bibnamefont {Gilligan}},\ }\href@noop {} {\bibfield
  {journal} {\bibinfo  {journal} {PLoS Comput. Biol.}\ }\textbf {\bibinfo
  {volume} {7}},\ \bibinfo {pages} {e1002174} (\bibinfo {year}
  {2011}{\natexlab{b}})}\BibitemShut {NoStop}%
\bibitem [{\citenamefont {Morone}\ and\ \citenamefont
  {Makse}(2015)}]{Morone_Nature2015}%
  \BibitemOpen
  \bibfield  {author} {\bibinfo {author} {\bibfnamefont {F.}~\bibnamefont
  {Morone}}\ and\ \bibinfo {author} {\bibfnamefont {H.~A.}\ \bibnamefont
  {Makse}},\ }\href {\doibase 10.1038/nature14604} {\bibfield  {journal}
  {\bibinfo  {journal} {Nature}\ }\textbf {\bibinfo {volume} {524}},\ \bibinfo
  {pages} {65} (\bibinfo {year} {2015})}\BibitemShut {NoStop}%
\bibitem [{\citenamefont {Habiba}\ \emph {et~al.}(2010)\citenamefont {Habiba},
  \citenamefont {Yu}, \citenamefont {Berger-Wolf},\ and\ \citenamefont
  {Saia}}]{Habiba_Chapter2010}%
  \BibitemOpen
  \bibfield  {author} {\bibinfo {author} {\bibnamefont {Habiba}}, \bibinfo
  {author} {\bibfnamefont {Y.}~\bibnamefont {Yu}}, \bibinfo {author}
  {\bibfnamefont {T.~Y.}\ \bibnamefont {Berger-Wolf}}, \ and\ \bibinfo {author}
  {\bibfnamefont {J.}~\bibnamefont {Saia}},\ }in\ \href {\doibase
  10.1007/978-3-642-14929-0_4} {\emph {\bibinfo {booktitle} {Adv. Soc. Netw.
  Min. Anal.}}}\ (\bibinfo  {publisher} {Springer},\ \bibinfo {address} {Berlin
  Heidelberg},\ \bibinfo {year} {2010})\ pp.\ \bibinfo {pages}
  {55--76}\BibitemShut {NoStop}%
\bibitem [{\citenamefont {Pei}\ and\ \citenamefont
  {Makse}(2013)}]{Pei_JSTAT2013}%
  \BibitemOpen
  \bibfield  {author} {\bibinfo {author} {\bibfnamefont {S.}~\bibnamefont
  {Pei}}\ and\ \bibinfo {author} {\bibfnamefont {H.~A.}\ \bibnamefont
  {Makse}},\ }\href {\doibase 10.1088/1742-5468/2013/12/P12002} {\bibfield
  {journal} {\bibinfo  {journal} {J. Stat. Mech. Theory Exp.}\ }\textbf
  {\bibinfo {volume} {2013}},\ \bibinfo {pages} {P12002} (\bibinfo {year}
  {2013})}\BibitemShut {NoStop}%
\bibitem [{\citenamefont {Wu}\ \emph {et~al.}(2015)\citenamefont {Wu},
  \citenamefont {Fu}, \citenamefont {Jin},\ and\ \citenamefont
  {Small}}]{Wu_PhysicaA2015_DynamicImmunization}%
  \BibitemOpen
  \bibfield  {author} {\bibinfo {author} {\bibfnamefont {Q.}~\bibnamefont
  {Wu}}, \bibinfo {author} {\bibfnamefont {X.}~\bibnamefont {Fu}}, \bibinfo
  {author} {\bibfnamefont {Z.}~\bibnamefont {Jin}}, \ and\ \bibinfo {author}
  {\bibfnamefont {M.}~\bibnamefont {Small}},\ }\href {\doibase
  10.1016/j.physa.2014.10.033} {\bibfield  {journal} {\bibinfo  {journal}
  {Physica A}\ }\textbf {\bibinfo {volume} {419}},\ \bibinfo {pages} {566}
  (\bibinfo {year} {2015})}\BibitemShut {NoStop}%
\bibitem [{\citenamefont {H{\'{e}}bert-Dufresne}\ \emph
  {et~al.}(2013)\citenamefont {H{\'{e}}bert-Dufresne}, \citenamefont {Allard},
  \citenamefont {Young},\ and\ \citenamefont
  {Dub{\'{e}}}}]{Hebert-Dufresne_SciRep2013}%
  \BibitemOpen
  \bibfield  {author} {\bibinfo {author} {\bibfnamefont {L.}~\bibnamefont
  {H{\'{e}}bert-Dufresne}}, \bibinfo {author} {\bibfnamefont {A.}~\bibnamefont
  {Allard}}, \bibinfo {author} {\bibfnamefont {J.-G.}\ \bibnamefont {Young}}, \
  and\ \bibinfo {author} {\bibfnamefont {L.~J.}\ \bibnamefont {Dub{\'{e}}}},\
  }\href {\doibase 10.1038/srep02171} {\bibfield  {journal} {\bibinfo
  {journal} {Sci. Rep.}\ }\textbf {\bibinfo {volume} {3}},\ \bibinfo {pages}
  {2171} (\bibinfo {year} {2013})}\BibitemShut {NoStop}%
\bibitem [{\citenamefont {Schneider}\ \emph {et~al.}(2012)\citenamefont
  {Schneider}, \citenamefont {Mihaljev},\ and\ \citenamefont
  {Herrmann}}]{Schneider_EPL2012}%
  \BibitemOpen
  \bibfield  {author} {\bibinfo {author} {\bibfnamefont {C.~M.}\ \bibnamefont
  {Schneider}}, \bibinfo {author} {\bibfnamefont {T.}~\bibnamefont {Mihaljev}},
  \ and\ \bibinfo {author} {\bibfnamefont {H.~J.}\ \bibnamefont {Herrmann}},\
  }\href {\doibase 10.1209/0295-5075/98/46002} {\bibfield  {journal} {\bibinfo
  {journal} {Europhys. Lett.}\ }\textbf {\bibinfo {volume} {98}},\ \bibinfo
  {pages} {46002} (\bibinfo {year} {2012})}\BibitemShut {NoStop}%
\bibitem [{\citenamefont {Kempe}\ \emph {et~al.}(2003)\citenamefont {Kempe},
  \citenamefont {Kleinberg},\ and\ \citenamefont {Tardos}}]{Kempe_2003}%
  \BibitemOpen
  \bibfield  {author} {\bibinfo {author} {\bibfnamefont {D.}~\bibnamefont
  {Kempe}}, \bibinfo {author} {\bibfnamefont {J.}~\bibnamefont {Kleinberg}}, \
  and\ \bibinfo {author} {\bibfnamefont {{\'{E}}.}~\bibnamefont {Tardos}},\
  }in\ \href {\doibase 10.1145/956750.956769} {\emph {\bibinfo {booktitle}
  {Proc. ninth ACM SIGKDD Int. Conf. Knowl. Discov. data Min. - KDD '03}}}\
  (\bibinfo  {publisher} {ACM Press},\ \bibinfo {address} {New York, New York,
  USA},\ \bibinfo {year} {2003})\ p.\ \bibinfo {pages} {137}\BibitemShut
  {NoStop}%
\bibitem [{\citenamefont {Liu}\ \emph {et~al.}(2016)\citenamefont {Liu},
  \citenamefont {Deng},\ and\ \citenamefont {Wei}}]{Liu_Chaos2016}%
  \BibitemOpen
  \bibfield  {author} {\bibinfo {author} {\bibfnamefont {Y.}~\bibnamefont
  {Liu}}, \bibinfo {author} {\bibfnamefont {Y.}~\bibnamefont {Deng}}, \ and\
  \bibinfo {author} {\bibfnamefont {B.}~\bibnamefont {Wei}},\ }\href {\doibase
  10.1063/1.4940240} {\bibfield  {journal} {\bibinfo  {journal} {Chaos}\
  }\textbf {\bibinfo {volume} {26}},\ \bibinfo {pages} {013106} (\bibinfo
  {year} {2016})}\BibitemShut {NoStop}%
\bibitem [{\citenamefont {Holme}(2004)}]{Holme_EPL2004_LocalVaccination}%
  \BibitemOpen
  \bibfield  {author} {\bibinfo {author} {\bibfnamefont {P.}~\bibnamefont
  {Holme}},\ }\href {\doibase 10.1209/epl/i2004-10286-2} {\bibfield  {journal}
  {\bibinfo  {journal} {Europhys. Lett.}\ }\textbf {\bibinfo {volume} {68}},\
  \bibinfo {pages} {908} (\bibinfo {year} {2004})}\BibitemShut {NoStop}%
\bibitem [{\citenamefont {Holme}\ \emph {et~al.}(2002)\citenamefont {Holme},
  \citenamefont {Kim}, \citenamefont {Yoon},\ and\ \citenamefont
  {Han}}]{Holme_PRE2002_AttackVulnerability}%
  \BibitemOpen
  \bibfield  {author} {\bibinfo {author} {\bibfnamefont {P.}~\bibnamefont
  {Holme}}, \bibinfo {author} {\bibfnamefont {B.~J.}\ \bibnamefont {Kim}},
  \bibinfo {author} {\bibfnamefont {C.~N.}\ \bibnamefont {Yoon}}, \ and\
  \bibinfo {author} {\bibfnamefont {S.~K.}\ \bibnamefont {Han}},\ }\href@noop
  {} {\bibfield  {journal} {\bibinfo  {journal} {Phys. Rev. E}\ }\textbf
  {\bibinfo {volume} {65}},\ \bibinfo {pages} {56109} (\bibinfo {year}
  {2002})}\BibitemShut {NoStop}%
\bibitem [{\citenamefont {Achlioptas}\ \emph {et~al.}(2009)\citenamefont
  {Achlioptas}, \citenamefont {D'Souza},\ and\ \citenamefont
  {Spencer}}]{Achlioptas_Science2009}%
  \BibitemOpen
  \bibfield  {author} {\bibinfo {author} {\bibfnamefont {D.}~\bibnamefont
  {Achlioptas}}, \bibinfo {author} {\bibfnamefont {R.~M.}\ \bibnamefont
  {D'Souza}}, \ and\ \bibinfo {author} {\bibfnamefont {J.}~\bibnamefont
  {Spencer}},\ }\href {\doibase 10.1126/science.1167782} {\bibfield  {journal}
  {\bibinfo  {journal} {Science}\ }\textbf {\bibinfo {volume} {323}},\ \bibinfo
  {pages} {1453} (\bibinfo {year} {2009})}\BibitemShut {NoStop}%
\bibitem [{Note1()}]{Note1}%
  \BibitemOpen
  \bibinfo {note} {At variance with Ref.~\cite {Achlioptas_Science2009}, where
  {\protect \it bond} percolation has been explored, here, it is more natural
  to deal with {\protect \it site} percolation: this is, however, a minor
  difference.}\BibitemShut {Stop}%
\bibitem [{\citenamefont
  {Saberi}(2015)}]{Saberi_PhysRep2015_ReviewPercolation}%
  \BibitemOpen
  \bibfield  {author} {\bibinfo {author} {\bibfnamefont {A.~A.}\ \bibnamefont
  {Saberi}},\ }\href {\doibase 10.1016/j.physrep.2015.03.003} {\bibfield
  {journal} {\bibinfo  {journal} {Phys. Rep.}\ }\textbf {\bibinfo {volume}
  {578}},\ \bibinfo {pages} {1} (\bibinfo {year} {2015})}\BibitemShut {NoStop}%
\bibitem [{\citenamefont {G{\'{o}}mez-Garde{\~{n}}es}\ \emph
  {et~al.}(2016)\citenamefont {G{\'{o}}mez-Garde{\~{n}}es}, \citenamefont
  {Lotero}, \citenamefont {Taraskin},\ and\ \citenamefont
  {P{\'{e}}rez-Reche}}]{GomezGardenes_PerezReche_SciRep2016}%
  \BibitemOpen
  \bibfield  {author} {\bibinfo {author} {\bibfnamefont {J.}~\bibnamefont
  {G{\'{o}}mez-Garde{\~{n}}es}}, \bibinfo {author} {\bibfnamefont
  {L.}~\bibnamefont {Lotero}}, \bibinfo {author} {\bibfnamefont {S.~N.}\
  \bibnamefont {Taraskin}}, \ and\ \bibinfo {author} {\bibfnamefont {F.~J.}\
  \bibnamefont {P{\'{e}}rez-Reche}},\ }\href {\doibase 10.1038/srep19767}
  {\bibfield  {journal} {\bibinfo  {journal} {Sci. Rep.}\ }\textbf {\bibinfo
  {volume} {6}},\ \bibinfo {pages} {19767} (\bibinfo {year}
  {2016})}\BibitemShut {NoStop}%
\bibitem [{\citenamefont {Janssen}\ \emph {et~al.}(2004)\citenamefont
  {Janssen}, \citenamefont {M{\"{u}}ller},\ and\ \citenamefont
  {Stenull}}]{Janssen_PRE2004}%
  \BibitemOpen
  \bibfield  {author} {\bibinfo {author} {\bibfnamefont {H.-K.}\ \bibnamefont
  {Janssen}}, \bibinfo {author} {\bibfnamefont {M.}~\bibnamefont
  {M{\"{u}}ller}}, \ and\ \bibinfo {author} {\bibfnamefont {O.}~\bibnamefont
  {Stenull}},\ }\href {\doibase 10.1103/PhysRevE.70.026114} {\bibfield
  {journal} {\bibinfo  {journal} {Phys. Rev. E}\ }\textbf {\bibinfo {volume}
  {70}},\ \bibinfo {pages} {026114} (\bibinfo {year} {2004})}\BibitemShut
  {NoStop}%
\bibitem [{\citenamefont {Bizhani}\ \emph {et~al.}(2012)\citenamefont
  {Bizhani}, \citenamefont {Paczuski},\ and\ \citenamefont
  {Grassberger}}]{Bizhani_PRE2012}%
  \BibitemOpen
  \bibfield  {author} {\bibinfo {author} {\bibfnamefont {G.}~\bibnamefont
  {Bizhani}}, \bibinfo {author} {\bibfnamefont {M.}~\bibnamefont {Paczuski}}, \
  and\ \bibinfo {author} {\bibfnamefont {P.}~\bibnamefont {Grassberger}},\
  }\href {\doibase 10.1103/PhysRevE.86.011128} {\bibfield  {journal} {\bibinfo
  {journal} {Phys. Rev. E}\ }\textbf {\bibinfo {volume} {86}},\ \bibinfo
  {pages} {11128} (\bibinfo {year} {2012})}\BibitemShut {NoStop}%
\bibitem [{\citenamefont {Chung}\ \emph {et~al.}(2014)\citenamefont {Chung},
  \citenamefont {Baek}, \citenamefont {Kim}, \citenamefont {Ha},\ and\
  \citenamefont {Jeong}}]{Chung_PRE2014}%
  \BibitemOpen
  \bibfield  {author} {\bibinfo {author} {\bibfnamefont {K.}~\bibnamefont
  {Chung}}, \bibinfo {author} {\bibfnamefont {Y.}~\bibnamefont {Baek}},
  \bibinfo {author} {\bibfnamefont {D.}~\bibnamefont {Kim}}, \bibinfo {author}
  {\bibfnamefont {M.}~\bibnamefont {Ha}}, \ and\ \bibinfo {author}
  {\bibfnamefont {H.}~\bibnamefont {Jeong}},\ }\href {\doibase
  10.1103/PhysRevE.89.052811} {\bibfield  {journal} {\bibinfo  {journal} {Phys.
  Rev. E}\ }\textbf {\bibinfo {volume} {89}},\ \bibinfo {pages} {052811}
  (\bibinfo {year} {2014})}\BibitemShut {NoStop}%
\bibitem [{\citenamefont {Dorogovtsev}\ \emph {et~al.}(2006)\citenamefont
  {Dorogovtsev}, \citenamefont {Goltsev},\ and\ \citenamefont
  {Mendes}}]{dorogovtsev2006k}%
  \BibitemOpen
  \bibfield  {author} {\bibinfo {author} {\bibfnamefont {S.~N.}\ \bibnamefont
  {Dorogovtsev}}, \bibinfo {author} {\bibfnamefont {A.~V.}\ \bibnamefont
  {Goltsev}}, \ and\ \bibinfo {author} {\bibfnamefont {J.~F.~F.}\ \bibnamefont
  {Mendes}},\ }\href@noop {} {\bibfield  {journal} {\bibinfo  {journal}
  {Physical review letters}\ }\textbf {\bibinfo {volume} {96}},\ \bibinfo
  {pages} {040601} (\bibinfo {year} {2006})}\BibitemShut {NoStop}%
\bibitem [{\citenamefont {Buldyrev}\ \emph {et~al.}(2010)\citenamefont
  {Buldyrev}, \citenamefont {Parshani}, \citenamefont {Paul}, \citenamefont
  {Stanley},\ and\ \citenamefont {Havlin}}]{Buldyrev_Nature2010}%
  \BibitemOpen
  \bibfield  {author} {\bibinfo {author} {\bibfnamefont {S.~V.}\ \bibnamefont
  {Buldyrev}}, \bibinfo {author} {\bibfnamefont {R.}~\bibnamefont {Parshani}},
  \bibinfo {author} {\bibfnamefont {G.}~\bibnamefont {Paul}}, \bibinfo {author}
  {\bibfnamefont {H.~E.}\ \bibnamefont {Stanley}}, \ and\ \bibinfo {author}
  {\bibfnamefont {S.}~\bibnamefont {Havlin}},\ }\href
  {http://dx.doi.org/10.1038/nature08932} {\bibfield  {journal} {\bibinfo
  {journal} {Nature}\ }\textbf {\bibinfo {volume} {464}},\ \bibinfo {pages}
  {1025} (\bibinfo {year} {2010})}\BibitemShut {NoStop}%
\bibitem [{\citenamefont {Son}\ \emph {et~al.}(2012)\citenamefont {Son},
  \citenamefont {Bizhani}, \citenamefont {Christensen}, \citenamefont
  {Grassberger},\ and\ \citenamefont {Paczuski}}]{Son-Grassberger_EPL2012}%
  \BibitemOpen
  \bibfield  {author} {\bibinfo {author} {\bibfnamefont {S.-W.}\ \bibnamefont
  {Son}}, \bibinfo {author} {\bibfnamefont {G.}~\bibnamefont {Bizhani}},
  \bibinfo {author} {\bibfnamefont {C.}~\bibnamefont {Christensen}}, \bibinfo
  {author} {\bibfnamefont {P.}~\bibnamefont {Grassberger}}, \ and\ \bibinfo
  {author} {\bibfnamefont {M.}~\bibnamefont {Paczuski}},\ }\href
  {http://stacks.iop.org/0295-5075/97/i=1/a=16006} {\bibfield  {journal}
  {\bibinfo  {journal} {Europhysics Lett.}\ }\textbf {\bibinfo {volume} {97}},\
  \bibinfo {pages} {16006} (\bibinfo {year} {2012})}\BibitemShut {NoStop}%
\bibitem [{\citenamefont {Gomez-Gardenes}\ \emph {et~al.}(2011)\citenamefont
  {Gomez-Gardenes}, \citenamefont {Gomez}, \citenamefont {Arenas},\ and\
  \citenamefont {Moreno}}]{Gardenes:2011}%
  \BibitemOpen
  \bibfield  {author} {\bibinfo {author} {\bibfnamefont {J.}~\bibnamefont
  {Gomez-Gardenes}}, \bibinfo {author} {\bibfnamefont {S.}~\bibnamefont
  {Gomez}}, \bibinfo {author} {\bibfnamefont {A.}~\bibnamefont {Arenas}}, \
  and\ \bibinfo {author} {\bibfnamefont {Y.}~\bibnamefont {Moreno}},\
  }\href@noop {} {\bibfield  {journal} {\bibinfo  {journal} {Phys. Rev. Lett.}\
  }\textbf {\bibinfo {volume} {106}},\ \bibinfo {pages} {128701} (\bibinfo
  {year} {2011})}\BibitemShut {NoStop}%
\bibitem [{\citenamefont {Leyva}\ \emph {et~al.}(2012)\citenamefont {Leyva},
  \citenamefont {Sevilla-Escoboza}, \citenamefont {Buldu}, \citenamefont
  {Sendina-Nadal}, \citenamefont {Gomez-Gardenes}, \citenamefont {Arenas},
  \citenamefont {Moreno}, \citenamefont {Gomez}, \citenamefont
  {Jaimes-Reategui},\ and\ \citenamefont {Boccaletti}}]{Leyva:2012}%
  \BibitemOpen
  \bibfield  {author} {\bibinfo {author} {\bibfnamefont {I.}~\bibnamefont
  {Leyva}}, \bibinfo {author} {\bibfnamefont {R.}~\bibnamefont
  {Sevilla-Escoboza}}, \bibinfo {author} {\bibfnamefont {J.~M.}\ \bibnamefont
  {Buldu}}, \bibinfo {author} {\bibfnamefont {I.}~\bibnamefont
  {Sendina-Nadal}}, \bibinfo {author} {\bibfnamefont {J.}~\bibnamefont
  {Gomez-Gardenes}}, \bibinfo {author} {\bibfnamefont {A.}~\bibnamefont
  {Arenas}}, \bibinfo {author} {\bibfnamefont {Y.}~\bibnamefont {Moreno}},
  \bibinfo {author} {\bibfnamefont {S.}~\bibnamefont {Gomez}}, \bibinfo
  {author} {\bibfnamefont {R.}~\bibnamefont {Jaimes-Reategui}}, \ and\ \bibinfo
  {author} {\bibfnamefont {S.}~\bibnamefont {Boccaletti}},\ }\href@noop {}
  {\bibfield  {journal} {\bibinfo  {journal} {Phys. Rev. Lett.}\ }\textbf
  {\bibinfo {volume} {108}},\ \bibinfo {pages} {168702} (\bibinfo {year}
  {2012})}\BibitemShut {NoStop}%
\bibitem [{\citenamefont {Motter}\ \emph {et~al.}(2013)\citenamefont {Motter},
  \citenamefont {Myers}, \citenamefont {Anghel},\ and\ \citenamefont
  {Nishikawa}}]{Motter:2013}%
  \BibitemOpen
  \bibfield  {author} {\bibinfo {author} {\bibfnamefont {A.~E.}\ \bibnamefont
  {Motter}}, \bibinfo {author} {\bibfnamefont {S.~A.}\ \bibnamefont {Myers}},
  \bibinfo {author} {\bibfnamefont {M.}~\bibnamefont {Anghel}}, \ and\ \bibinfo
  {author} {\bibfnamefont {T.}~\bibnamefont {Nishikawa}},\ }\href@noop {}
  {\bibfield  {journal} {\bibinfo  {journal} {Nat. Phys.}\ }\textbf {\bibinfo
  {volume} {9}},\ \bibinfo {pages} {191} (\bibinfo {year} {2013})}\BibitemShut
  {NoStop}%
\bibitem [{\citenamefont {Ji}\ \emph {et~al.}(2013)\citenamefont {Ji},
  \citenamefont {Peron}, \citenamefont {Menck}, \citenamefont {Rodrigues},\
  and\ \citenamefont {Kurths}}]{Kurths:2013}%
  \BibitemOpen
  \bibfield  {author} {\bibinfo {author} {\bibfnamefont {P.}~\bibnamefont
  {Ji}}, \bibinfo {author} {\bibfnamefont {T.~K.~D.}\ \bibnamefont {Peron}},
  \bibinfo {author} {\bibfnamefont {P.~J.}\ \bibnamefont {Menck}}, \bibinfo
  {author} {\bibfnamefont {F.~A.}\ \bibnamefont {Rodrigues}}, \ and\ \bibinfo
  {author} {\bibfnamefont {J.}~\bibnamefont {Kurths}},\ }\href@noop {}
  {\bibfield  {journal} {\bibinfo  {journal} {Phys. Rev. Lett.}\ }\textbf
  {\bibinfo {volume} {110}},\ \bibinfo {pages} {218701} (\bibinfo {year}
  {2013})}\BibitemShut {NoStop}%
\bibitem [{\citenamefont {Echenique}\ \emph {et~al.}(2005)\citenamefont
  {Echenique}, \citenamefont {G{\'{o}}mez-Garde{\~{n}}es},\ and\ \citenamefont
  {Moreno}}]{Echenique2005}%
  \BibitemOpen
  \bibfield  {author} {\bibinfo {author} {\bibfnamefont {P.}~\bibnamefont
  {Echenique}}, \bibinfo {author} {\bibfnamefont {J.}~\bibnamefont
  {G{\'{o}}mez-Garde{\~{n}}es}}, \ and\ \bibinfo {author} {\bibfnamefont
  {Y.}~\bibnamefont {Moreno}},\ }\href {\doibase 10.1209/epl/i2005-10080-8}
  {\bibfield  {journal} {\bibinfo  {journal} {Europhys. Lett.}\ }\textbf
  {\bibinfo {volume} {71}},\ \bibinfo {pages} {325} (\bibinfo {year}
  {2005})}\BibitemShut {NoStop}%
\bibitem [{\citenamefont {Newman}\ and\ \citenamefont
  {Ziff}(2001)}]{Newman-Ziff_PRE2001}%
  \BibitemOpen
  \bibfield  {author} {\bibinfo {author} {\bibfnamefont {M.~E.~J.}\
  \bibnamefont {Newman}}\ and\ \bibinfo {author} {\bibfnamefont {R.~M.}\
  \bibnamefont {Ziff}},\ }\href {\doibase 10.1103/PhysRevE.64.016706}
  {\bibfield  {journal} {\bibinfo  {journal} {Phys. Rev. E}\ }\textbf {\bibinfo
  {volume} {64}},\ \bibinfo {pages} {016706} (\bibinfo {year}
  {2001})}\BibitemShut {NoStop}%
\bibitem [{\citenamefont {Braunstein}\ \emph {et~al.}(2016)\citenamefont
  {Braunstein}, \citenamefont {Dall'Asta}, \citenamefont {Semerjian},\ and\
  \citenamefont {Zdeborov{\'a}}}]{braunstein2016network}%
  \BibitemOpen
  \bibfield  {author} {\bibinfo {author} {\bibfnamefont {A.}~\bibnamefont
  {Braunstein}}, \bibinfo {author} {\bibfnamefont {L.}~\bibnamefont
  {Dall'Asta}}, \bibinfo {author} {\bibfnamefont {G.}~\bibnamefont
  {Semerjian}}, \ and\ \bibinfo {author} {\bibfnamefont {L.}~\bibnamefont
  {Zdeborov{\'a}}},\ }\href@noop {} {\bibfield  {journal} {\bibinfo  {journal}
  {arXiv preprint arXiv:1603.08883}\ } (\bibinfo {year} {2016})}\BibitemShut
  {NoStop}%
\bibitem [{\citenamefont {Mugisha}\ and\ \citenamefont
  {Zhou}(2016)}]{mugisha2016identifying}%
  \BibitemOpen
  \bibfield  {author} {\bibinfo {author} {\bibfnamefont {S.}~\bibnamefont
  {Mugisha}}\ and\ \bibinfo {author} {\bibfnamefont {H.-J.}\ \bibnamefont
  {Zhou}},\ }\href@noop {} {\bibfield  {journal} {\bibinfo  {journal} {arXiv
  preprint arXiv:1603.05781}\ } (\bibinfo {year} {2016})}\BibitemShut {NoStop}%
\bibitem [{Note2()}]{Note2}%
  \BibitemOpen
  \bibinfo {note} {Notice that $m=2$ was used in \cite
  {Achlioptas_Science2009}, in the context of EP.}\BibitemShut {Stop}%
\bibitem [{\citenamefont {Kov{\'{a}}cs}\ and\ \citenamefont
  {Barab{\'{a}}si}(2015)}]{Kovacs-Barabasi_Nature2015}%
  \BibitemOpen
  \bibfield  {author} {\bibinfo {author} {\bibfnamefont {I.~A.}\ \bibnamefont
  {Kov{\'{a}}cs}}\ and\ \bibinfo {author} {\bibfnamefont {A.-L.}\ \bibnamefont
  {Barab{\'{a}}si}},\ }\href {\doibase 10.1038/524038a} {\bibfield  {journal}
  {\bibinfo  {journal} {Nature}\ }\textbf {\bibinfo {volume} {524}},\ \bibinfo
  {pages} {38} (\bibinfo {year} {2015})}\BibitemShut {NoStop}%
\bibitem [{\citenamefont {Boettcher}\ and\ \citenamefont
  {Percus}(2001)}]{Boettcher-Percus_PRL2001}%
  \BibitemOpen
  \bibfield  {author} {\bibinfo {author} {\bibfnamefont {S.}~\bibnamefont
  {Boettcher}}\ and\ \bibinfo {author} {\bibfnamefont {A.~G.}\ \bibnamefont
  {Percus}},\ }\href {\doibase 10.1103/PhysRevLett.86.5211} {\bibfield
  {journal} {\bibinfo  {journal} {Phys. Rev. Lett.}\ }\textbf {\bibinfo
  {volume} {86}},\ \bibinfo {pages} {5211} (\bibinfo {year}
  {2001})}\BibitemShut {NoStop}%
\bibitem [{\citenamefont {Cohen}\ \emph {et~al.}(2001)\citenamefont {Cohen},
  \citenamefont {Erez}, \citenamefont {Ben-Avraham},\ and\ \citenamefont
  {Havlin}}]{Cohen_PRL2001}%
  \BibitemOpen
  \bibfield  {author} {\bibinfo {author} {\bibfnamefont {R.}~\bibnamefont
  {Cohen}}, \bibinfo {author} {\bibfnamefont {K.}~\bibnamefont {Erez}},
  \bibinfo {author} {\bibfnamefont {D.}~\bibnamefont {Ben-Avraham}}, \ and\
  \bibinfo {author} {\bibfnamefont {S.}~\bibnamefont {Havlin}},\ }\href
  {\doibase 10.1103/PhysRevLett.86.3682} {\bibfield  {journal} {\bibinfo
  {journal} {Phys. Rev. Lett.}\ }\textbf {\bibinfo {volume} {86}},\ \bibinfo
  {pages} {3682} (\bibinfo {year} {2001})}\BibitemShut {NoStop}%
\bibitem [{\citenamefont {Barab{\'{a}}si}\ and\ \citenamefont
  {Albert}(1999)}]{Barabasi-Albert_Science1999}%
  \BibitemOpen
  \bibfield  {author} {\bibinfo {author} {\bibfnamefont {A.-L.}\ \bibnamefont
  {Barab{\'{a}}si}}\ and\ \bibinfo {author} {\bibfnamefont {R.}~\bibnamefont
  {Albert}},\ }\href {\doibase 10.1126/science.286.5439.509} {\bibfield
  {journal} {\bibinfo  {journal} {Science (80-. ).}\ }\textbf {\bibinfo
  {volume} {286}},\ \bibinfo {pages} {11} (\bibinfo {year} {1999})}\BibitemShut
  {NoStop}%
\bibitem [{\citenamefont {Keeling}\ \emph {et~al.}(2003)\citenamefont
  {Keeling}, \citenamefont {Woolhouse}, \citenamefont {May}, \citenamefont
  {Davies},\ and\ \citenamefont {Grenfell}}]{Keeling_Nature2003}%
  \BibitemOpen
  \bibfield  {author} {\bibinfo {author} {\bibfnamefont {M.~J.}\ \bibnamefont
  {Keeling}}, \bibinfo {author} {\bibfnamefont {M.~E.~J.}\ \bibnamefont
  {Woolhouse}}, \bibinfo {author} {\bibfnamefont {R.~M.}\ \bibnamefont {May}},
  \bibinfo {author} {\bibfnamefont {G.}~\bibnamefont {Davies}}, \ and\ \bibinfo
  {author} {\bibfnamefont {B.~T.}\ \bibnamefont {Grenfell}},\ }\href
  {http://dx.doi.org/10.1038/nature01343} {\bibfield  {journal} {\bibinfo
  {journal} {Nature}\ }\textbf {\bibinfo {volume} {421}},\ \bibinfo {pages}
  {136} (\bibinfo {year} {2003})}\BibitemShut {NoStop}%
\bibitem [{\citenamefont {Kao}\ \emph {et~al.}(2007)\citenamefont {Kao},
  \citenamefont {Green}, \citenamefont {Johnson},\ and\ \citenamefont
  {Kiss}}]{Kao_JRoyInterface2007}%
  \BibitemOpen
  \bibfield  {author} {\bibinfo {author} {\bibfnamefont {R.~R.}\ \bibnamefont
  {Kao}}, \bibinfo {author} {\bibfnamefont {D.~M.}\ \bibnamefont {Green}},
  \bibinfo {author} {\bibfnamefont {J.}~\bibnamefont {Johnson}}, \ and\
  \bibinfo {author} {\bibfnamefont {I.~Z.}\ \bibnamefont {Kiss}},\ }\href
  {\doibase 10.1098/rsif.2007.1129} {\bibfield  {journal} {\bibinfo  {journal}
  {J. R. Soc. Interface}\ }\textbf {\bibinfo {volume} {4}},\ \bibinfo {pages}
  {907} (\bibinfo {year} {2007})}\BibitemShut {NoStop}%
\bibitem [{cat()}]{cattle_data}%
  \BibitemOpen
  \href@noop {} {}\bibinfo {note} {Data provided by AHVLA RADAR. The network
  can be downloaded from http://link.aps.org/supplemental/xxx}\BibitemShut
  {NoStop}%
\bibitem [{Note3()}]{Note3}%
  \BibitemOpen
  \bibinfo {note} {See
  http://vlado.fmf.uni-lj.si/pub/networks/data/hep-th/hep-th.htm}\BibitemShut
  {NoStop}%
\bibitem [{Note4()}]{Note4}%
  \BibitemOpen
  \bibinfo {note} {Data downloaded from http://www.netdimes.org for the year 2012.}\BibitemShut {NoStop}%
\bibitem [{\citenamefont {P{\'{e}}rez-Reche}\ \emph {et~al.}(2012)\citenamefont
  {P{\'{e}}rez-Reche}, \citenamefont {Taraskin}, \citenamefont {Otten},
  \citenamefont {Viana}, \citenamefont {Costa},\ and\ \citenamefont
  {Gilligan}}]{PerezReche_PRL2012_Soil}%
  \BibitemOpen
  \bibfield  {author} {\bibinfo {author} {\bibfnamefont {F.~J.}\ \bibnamefont
  {P{\'{e}}rez-Reche}}, \bibinfo {author} {\bibfnamefont {S.~N.}\ \bibnamefont
  {Taraskin}}, \bibinfo {author} {\bibfnamefont {W.}~\bibnamefont {Otten}},
  \bibinfo {author} {\bibfnamefont {M.~P.}\ \bibnamefont {Viana}}, \bibinfo
  {author} {\bibfnamefont {L.~D.~F.}\ \bibnamefont {Costa}}, \ and\ \bibinfo
  {author} {\bibfnamefont {C.~a.}\ \bibnamefont {Gilligan}},\ }\href {\doibase
  10.1103/PhysRevLett.109.098102} {\bibfield  {journal} {\bibinfo  {journal}
  {Phys. Rev. Lett.}\ }\textbf {\bibinfo {volume} {109}},\ \bibinfo {pages}
  {098102} (\bibinfo {year} {2012})}\BibitemShut {NoStop}%
\bibitem [{Air()}]{Airports_data}%
  \BibitemOpen
  \href@noop {} {}\bibinfo {note}
  {Data from the OpenFlights Airports Database (http://openflights.org/data.html) on 2009.}\BibitemShut {NoStop}%
\bibitem [{\citenamefont {Girvan}\ and\ \citenamefont
  {Newman}(2002)}]{Girvan-Newman_PNAS2002}%
  \BibitemOpen
  \bibfield  {author} {\bibinfo {author} {\bibfnamefont {M.}~\bibnamefont
  {Girvan}}\ and\ \bibinfo {author} {\bibfnamefont {M.~E.}\ \bibnamefont
  {Newman}},\ }\href
  {http://www.pnas.org/content/99/12/7821.short$\backslash$nhttp://www.ncbi.nlm.nih.gov/pmc/articles/PMC122977/}
  {\bibfield  {journal} {\bibinfo  {journal} {Proc. Natl. Acad. Sci.}\ }\textbf
  {\bibinfo {volume} {99}},\ \bibinfo {pages} {7821} (\bibinfo {year}
  {2002})}\BibitemShut {NoStop}%
\end{thebibliography}
\end{document}